\documentclass{aa}  
\usepackage[breaklinks=true,colorlinks,citecolor=blue]{hyperref}
\usepackage{graphicx}
\usepackage{enumerate}
\usepackage{epsfig}
\usepackage{color}
\usepackage{txfonts}
\usepackage{natbib}
\usepackage{multirow}
\usepackage{ulem} 
\usepackage{amssymb}
\usepackage{xcolor}

\definecolor{ao}{rgb}{0.0, 0.5, 0.0}

\defcitealias{KhoperskovHESTIA-1}{Paper I}
\defcitealias{KhoperskovHESTIA-2}{Paper II}
\defcitealias{KhoperskovHESTIA-3}{Paper III}

\newcommand{\kmps}{\rm km~s\ensuremath{^{-1} }\,}

\newcommand{\Msun}{M\ensuremath{_\odot}}

\newcommand{\Gaia}{{\it Gaia}\,}

\newcommand{\RVphi}{\rm ($R$,v\ensuremath{_\phi})\,}

\newcommand{\VV}{V\ensuremath{_\phi}\,}

\newcommand{\uv}{V\ensuremath{_R}-V\ensuremath{_\phi}\,}
\newcommand{\UV}{(V\ensuremath{_R},V\ensuremath{_\phi})\,}

\newcommand{\aFe}{\ensuremath{\rm [\alpha/Fe]}\,}

\newcommand{\FeH}{\ensuremath{\rm [Fe/H]}\,}

\newcommand{\Elz}{\ensuremath{\rm L_z-E}\,}
\newcommand{\ELz}{\ensuremath{\rm L_z-E}\,}
\newcommand{\JJtot}{\ensuremath{(J_r-J_z)/J_{tot} - J_\phi/J_{tot}}\,}
\newcommand{\sJrJp}{\ensuremath{\sqrt{J_r}-J_\phi}\,}

\begin{document} 

\title{The stellar halo in Local Group Hestia simulations II. \\ The accreted component}

\titlerunning{HESTIA simulations: phase-space composition of accreted halo}

\author{Sergey Khoperskov$^1$\thanks{E-mail: sergey.khoperskov@gmail.com}, Ivan Minchev$^1$, Noam Libeskind$^{1,2}$, Misha Haywood$^3$, Paola Di Matteo$^3$, \\ Vasily Belokurov$^{4,5}$, Matthias Steinmetz$^1$,  Facundo A. Gomez$^{6,7}$, Robert J. J. Grand$^{8,9,10}$,  Yehuda Hoffman$^{11}$, Alexander Knebe$^{12,13,14}$,  Jenny G. Sorce$^{15.16,1}$, Martin Spaare$^{17,1}$, Elmo Tempel$^{18,19}$, Mark Vogelsberger$^{20}$}

\authorrunning{Khoperskov et al.}

\institute{$^1$ Leibniz-Institut für Astrophysik Potsdam (AIP), An der Sternwarte 16, 14482 Potsdam, Germany\\
        $^2$ University of Lyon, UCB Lyon 1, CNRS/IN2P3, IUF, IP2I Lyon, France \\
        $^3$ GEPI, Observatoire de Paris, PSL Research University, CNRS, Place Jules Janssen, 92195 Meudon, France \\
        $^4$ Institute of Astronomy, Madingley Road, Cambridge CB3 0HA, UK \\
        $^5$ Center for Computational Astrophysics, Flatiron Institute, 162 5th Avenue, New York, NY 10010, USA \\
        $^6$ Instituto de Investigación Multidisciplinar en Ciencia y Tecnología, Universidad de La Serena, Raúl Bitrán 1305, La Serena, Chile \\
        $^7$ Departamento de Astronomía, Universidad de La Serena, Av. Juan Cisternas 1200 Norte, La Serena, Chile \\
        $^8$ Max-Planck-Institut für Astrophysik, Karl-Schwarzschild-Str 1, D-85748 Garching, Germany \\
        $^9$ Instituto de Astrofísica de Canarias, Calle Váa Láctea s/n, E-38205 La Laguna, Tenerife, Spain\\
        $^{10}$ Departamento de Astrofísica, Universidad de La Laguna, Av. del Astrofísico Francisco Sánchez s/n, E-38206 La Laguna, Tenerife, Spain\\
        $^{11}$ Racah Institute of Physics, Hebrew University, Jerusalem 91904, Israel \\
        $^{12}$ Departamento de Física Teórica, Módulo 15, Facultad de Ciencias, Universidad Autónoma de Madrid, E-28049 Madrid, Spain \\
        $^{13}$ Centro de Investigación Avanzada en Física Fundamental (CIAFF), Facultad de Ciencias, Universidad Autónoma de Madrid, E-28049 Madrid, Spain \\ 
        $^{14}$ International Centre for Radio Astronomy Research, University of Western Australia, 35 Stirling Highway, Crawley, Western Australia 6009, Australia \\
        $^{15}$ Univ. Lille, CNRS, Centrale Lille, UMR 9189 CRIStAL, F-59000 Lille, France\\
        $^{16}$ Universit\'e Paris-Saclay, CNRS, Institut d'Astrophysique Spatiale, 91405, Orsay, France\\
        $^{17}$ Institut für Physik und Astronomie, Universität Potsdam, Campus Golm, Haus 28, Karl-Liebknecht Straße 24-25, D-14476 Potsdam \\
        $^{18}$ Tartu Observatory, University of Tartu, Observatooriumi 1, 61602 Tõravere, Estonia \\
        $^{19}$ Estonian Academy of Sciences, Kohtu 6, 10130 Tallinn, Estonia \\
        $^{20}$ Department of Physics, Kavli Institute for Astrophysics and Space Research, Massachusetts Institute of Technology, Cambridge, MA 02139, USA
        }

\abstract{
 Recent progress in understanding the assembly history of the Milky Way~(MW) is driven by the tremendous amount of high-quality data delivered by \Gaia (ESA), revealing a number of substructures potentially linked to several ancient accretion events. In this work we aim to explore the phase-space structure of accreted stars by analysing six M31/MW analogues from the HESTIA suite of cosmological hydrodynamics zoom-in simulations of the Local Group. We find that all HESTIA galaxies experience a few dozen mergers but only between one and four of those have stellar mass ratios $>0.2$, relative to the host at the time of the merger. Depending on the halo definition, the most massive merger contributes from $20\%$ to $70\%$ of the total stellar halo mass. 
 Individual merger remnants show diverse density distributions at $\rm z=0$, significantly overlapping with each other and with the in situ stars in the \ELz, \UV and \RVphi coordinates. Moreover, merger debris often shifts position in the \ELz space with cosmic time due to the galactic mass growth and the non-axisymmetry of the potential. In agreement with previous works, we show that even individual merger debris exhibit a number of distinct \ELz features. In the \UV plane, all HESTIA galaxies reveal radially hot, non-rotating or weakly counter-rotating, Gaia-Sausage-like features, which are the remnants of the most recent significant mergers. We find an age gradient in \Elz space for individual debris, where the youngest stars, formed in the inner regions of accreting systems, deposit to the innermost regions of the host galaxies. The bulk of these stars formed during the last stages of accretion, making it possible to use the stellar ages of the remnants to date the merger event.
In action space~($J_r, J_z, J_\phi$), merger debris do not appear as isolated substructures, but are instead scattered over a large parameter area and overlap with the in situ stars. We suggest that accreted stars can be best identified using $\sqrt{J_r}>0.2-0.3\ {\rm (10^4\ kpc\ km\ s^{-1})^{0.5}}$. We also introduce a new, purely kinematic space ($J_z/J_r-$orbital eccentricity), where different merger debris can be disentangled better from each other and from the in situ stars. Accreted stars have a broad distribution of eccentricities, peaking at $\epsilon\approx0.6-0.9$, and their mean eccentricity tends to be smaller for systems accreted more recently. 
}


\maketitle

\section{Introduction}\label{sec2::intro}

Thanks to a long dynamical timescale, a stellar halo serves as a fossil record, revealing the galactic assembly history~\citep{1999Natur.402...53H, 2005ApJ...635..931B,2006MNRAS.365..747A}. Stars stripped off from different dwarf galaxies can be identified as streams or moving groups in the halo~\citep{1994Natur.370..194I,2001Natur.412...49I, 2007ApJ...658..337B}, which are visible on top of the in situ stellar halo formed via either dissipative collapse~\citep{1962ApJ...136..748E,1987AJ.....93...74S} or heating a pre-existing stellar disc~\citep{2009ApJ...702.1058Z,2010MNRAS.404.1711P}. Stars from accreted massive satellites result in large-scale stellar substructures in the halo, which are not spatially coherent at $\rm z=0$, and are thus expected to preserve  their common origin imprinted into their chemical and kinematical properties~\citep{2005ApJ...635..931B, 2006MNRAS.365..747A,2008ApJ...683..597S, 2008MNRAS.391...14D, 2010MNRAS.406..744C, 2011MNRAS.416.2802F, 2012MNRAS.426..690B, 2015ApJ...799..184P, 2016MNRAS.458.2371R,2016ApJ...821....5D}. 

Although merger stellar remnants disperse across a large volume in the halo, it is believed that even after a long time they can still be identified in purely kinematic, phase-space coordinates or in integrals of motion~(e.g., energy-angular momentum, \ELz). For instance, \cite{1999MNRAS.307..495H} showed that there is a strong correlation between different velocity components of merger debris, making it possible to associate certain kinematic groups with particular mergers~\citep[see, also, e.g.,][]{2009MNRAS.399..166V,2019A&A...631L...9K}. Broadly speaking, a number of models suggest that individual merger remnants can be identified as coherent structures in the total energy-angular momentum components space~\citep[\ELz, see, e.g.][]{1996ApJ...465..278J,2005A&A...439..551R,2007AJ....134.1579K, 2000MNRAS.319..657H,2007MNRAS.381..987C, 2009ApJ...694..130M, 2010MNRAS.408..935G,2010MNRAS.401.2285G,2015AJ....150..128R}. These results are based on the assumption that energy and angular momentum are both approximately conserved over the galaxy life-time and thus the merger debris stars follow trajectories of their progenitor systems~\citep[see Sec.3.3 in][for the caveats]{2020ARA&A..58..205H}. However, simulations with evolving galactic potentials suggest that the merger remnants can be smeared significantly in the \ELz coordinates due to dynamical friction, mass growth and halo shape evolution, and this finding is valid for the baryonic components~\citep{2005MNRAS.357L..35K,2017MNRAS.464.2882A,2017A&A...604A.106J,2018MNRAS.473.1656T,2019MNRAS.487L..72G,2021ApJ...920...10P,2021MNRAS.501.2279V,2022arXiv220412187A} and for dark matter \citep{2005MNRAS.357L..35K}. Moreover, the finite size of accreting dwarf galaxies and their own self-gravity will also affect the coherence of their merger debris~\citep[][]{2013MNRAS.433.1813S}. Recently, \cite{Pagnini2022} showed that the globular clusters of accreted dwarf galaxies barely trace their stellar debris progenitor in the kinematic space. It is also worth mentioning that chaotic mixing may affect the tidal streams' density on a rather short timescale, making it hard to isolate the debris, especially in a small volume of space~\citep[][]{2008MNRAS.385..236V,2016MNRAS.455.1079P}. However, \cite{2015MNRAS.453.2830M,2018MNRAS.478.4052M} showed that diffusion due to chaotic mixing in the neighbourhood of the Sun does not efficiently erase phase space signatures of past accretion events.

Nowadays, the Milky Way~(MW) is the best laboratory for studying the assembly history of a massive galaxy via the identification of merger debris, since the number of stars with 6D phase-space information has been boosted drastically by data from the \Gaia~(ESA) satellite~\citep{2016A&A...595A...2G,2018A&A...616A...1G}. These data complemented by the information from spectroscopic surveys make it possible to identify the merger remnants and disentangle them from the in situ disc and halo populations not only in kinematic space but also in the chemical abundances~\citep{2019MNRAS.482.3426M,2020MNRAS.493.3363H,2020ApJ...901...48N,2021SCPMA..6439562Z,2021A&A...650A.110M,2021MNRAS.507...43S}.

Prior to the \Gaia mission, \cite{2010A&A...511L..10N} and \cite{2012A&A...538A..21S} were the first to identify the presence of two distinct halo populations at  a metallicity lower than that of the Galactic disc in the solar neighbourhood~\citep[see, also,][]{2000AJ....119.2843C,2001ApJ...549..325C,2003ApJ...585L.125B}.  More recently, using \Gaia data, \cite{2018MNRAS.478..611B} discovered stellar debris in the inner halo deposited in a major accretion event by a satellite between $8$ and $11$ Gyr ago~\citep[Gaia-Sausage, see also,][]{2018ApJ...863L..28M, 2018ApJ...862L...1D}. The Gaia-Sausage populations is the most distinct in the \UV space at $-2<\FeH<-1$ where it corresponds to the non-rotating component with large radial orbital excursion. The second \Gaia data release revealed a double colour–magnitude sequence~\citep{2018A&A...616A..10G} observed for the stars with high transverse velocities which was used to  more closely constrain the last significant merger in  MW history~\citep[][Gaia-Enceladus]{2018ApJ...863..113H,2018Natur.563...85H} which, similar to \cite{2010A&A...511L..10N}, can be identified as a distinct low-\FeH sequence in \aFe-\FeH plane.

A number of recent works provide the estimates of the stellar mass of accreted Gaia-Sausage-Enceladus~(GSE) satellite to be in the range of $2\times10^{8}-5\times10^9$~\Msun~\citep{2018ApJ...852...50F,2019MNRAS.487L..47V,2019MNRAS.484.4471F,2019MNRAS.482.3426M,2020MNRAS.497..109F}. Theory suggests that such massive accretion events significantly perturb the pre-existing disc  populations~\citep{1993ApJ...403...74Q,1999MNRAS.304..254V,2016MNRAS.456.2779G,2016MNRAS.459..199G,2017MNRAS.472.3722G}. In the MW the impact of the GSE merger is associated with the heating of the in situ populations~\citep[``Splash'' in ][or ``Plume'' as it was introduced in \cite{2019A&A...632A...4D}]{2020MNRAS.494.3880B}. The stellar kinematic fossil record shows the imprint left by this accretion event, which heated the (metal-rich) thick disc which is now found in the Galactic halo leaving little room for the existence of in situ stellar halo component~\citep[][but see recent findings by \cite{2022arXiv220304980B}]{2019A&A...632A...4D}.

%%%%%%%%%%%%%%%%%%%%%%%%%%%%%%%%%%%%%%%%%%%%%%%%%%%%%%%%%
\begin{figure*}[t!]
\begin{center}
\includegraphics[width=0.8\hsize]{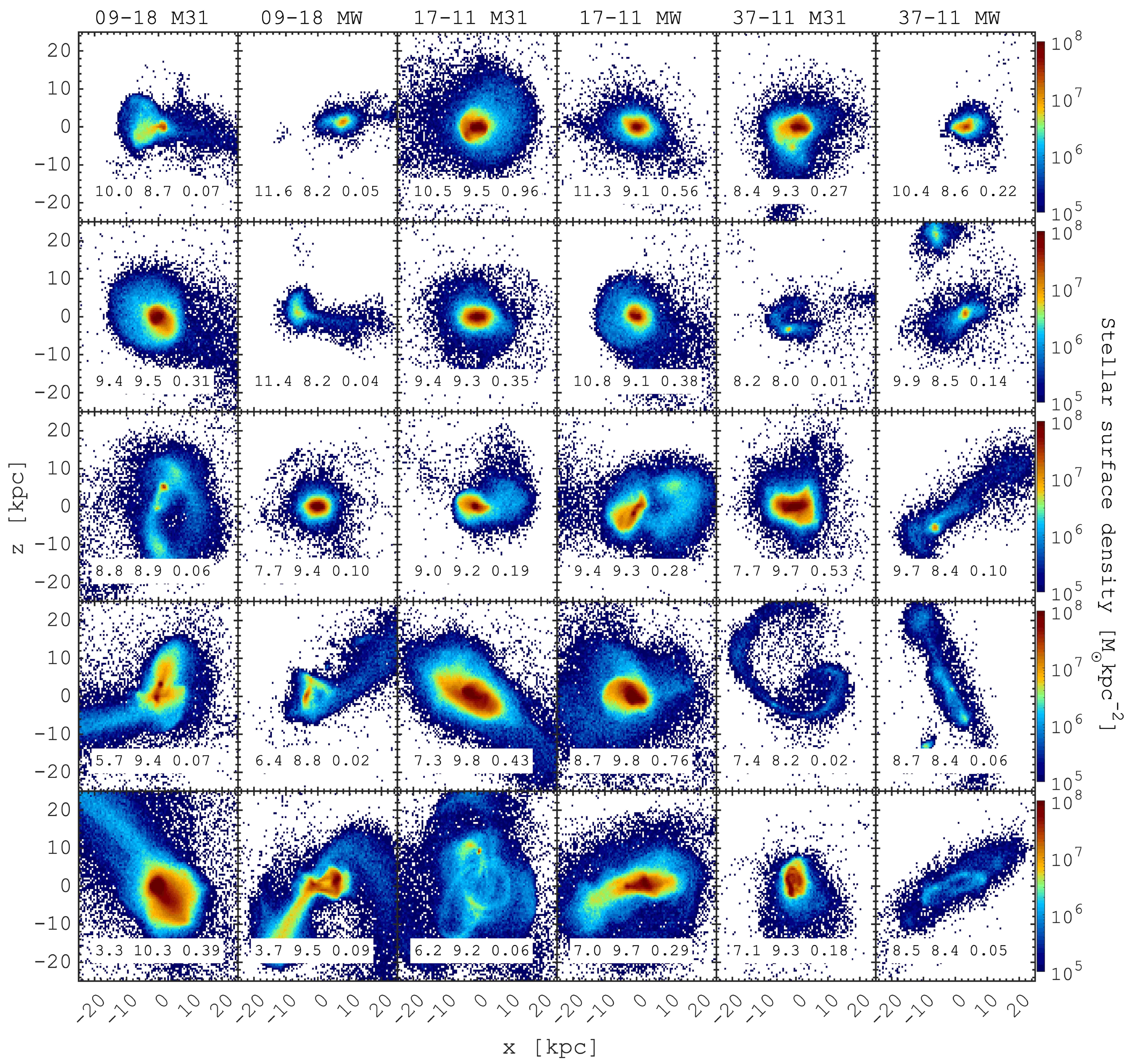}
\caption{Morphology of individual mergers at the time of accretion for the five most significant events, for each M31/MW analogues in HESTIA. The merger accretion lookback time~(Gyr), total stellar mass of the merger debris at the time of the merger ($\rm log_{10}(M_{*}/M_\odot)$), and the stellar mass ratio ($\mu_{*}$) relative to the main M31/MW progenitor at the time of the merger are given at the bottom of each panel. The colour bar represents stellar density, in units of $\rm M_{*}/kpc^2$. The merger debris show diverse morphologies, likely caused by the different trajectories of accreting systems and internal mass/kinematics distributions.}
\label{fig2::mergers_initial}
\end{center}
\end{figure*}
%%%%%%%%%%%%%%%%%%%%%%%%%%%%%%%%%%%%%%%%%%%%%%%%%%%%%%%%%

%%%%%%%%%%%%%%%%%%%%%%%%%%%%%%%%%%%%%%%%%%%%%%%%%%%%%%%%%
\begin{figure*}[t!]
\begin{center}
\includegraphics[width=0.8\hsize]{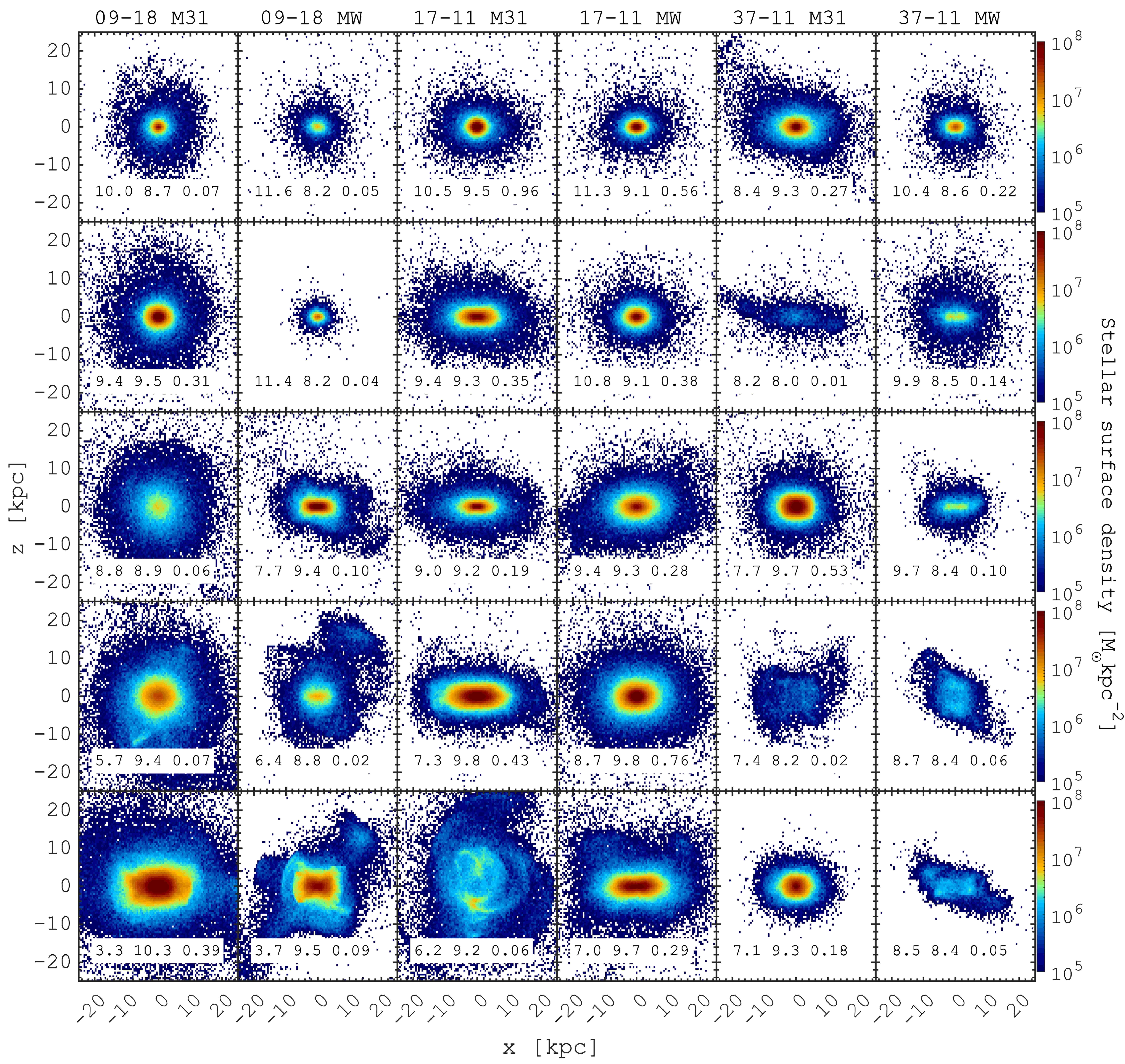}
\caption{Same as Fig.~\ref{fig2::mergers_initial}, but merger debris are shown at $\rm z=0$ instead of at the time of accretion. At this time they appear well phased-mixed with some structure found for the most recent events and the earlier mergers seem to be more compact compared to the later ones. }\label{fig2::mergers_final}
\end{center}
\end{figure*}
%%%%%%%%%%%%%%%%%%%%%%%%%%%%%%%%%%%%%%%%%%%%%%%%%%%%%%%%%

%%%%%%%%%%%%%%%%%%%%%%%%%%%%%%%%%%%%%%%%%%%%%%%%%%%%%%%%%
\begin{figure*}[t!]
\begin{center}
\includegraphics[width=0.99\hsize]{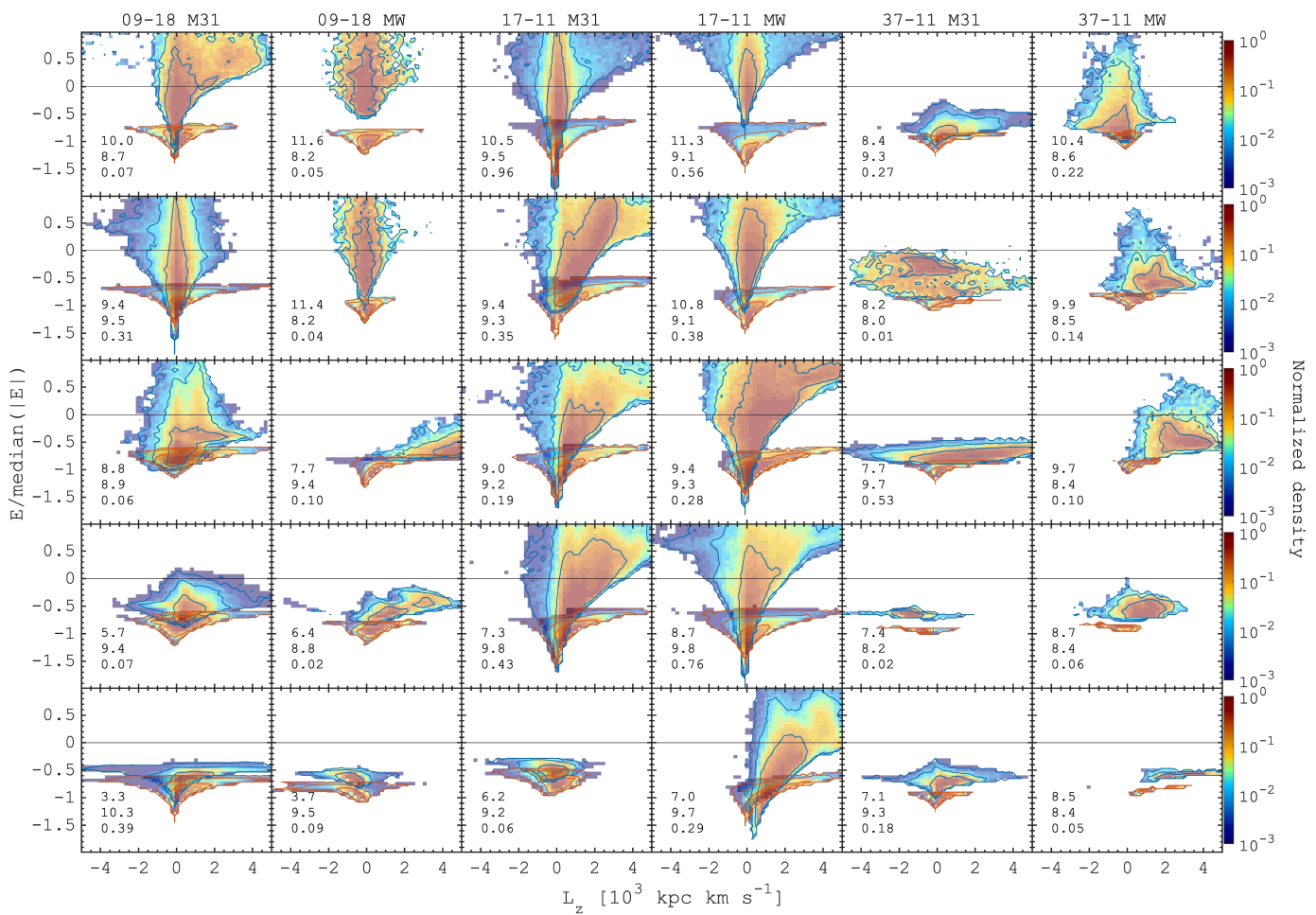}
\caption{Energy-angular momentum relation for the stellar remnants. Five most significant mergers~(M1-M5; see Fig.~4 in \citetalias{KhoperskovHESTIA-1}) are shown at the time of accretion~(blue contours) and with at $\rm z=0$~(red contours). 
The merger accretion lookback time~(Gyr), total stellar mass of the merger debris at the time of the merger ($\rm log_{10}(M_{*}/M_\odot)$), and the stellar mass ratio ($\mu_{*}$) relative to the main M31/MW progenitor at the time of the merger are given in the bottom left corner of each panel. Since the total host mass increases over time, the initial \ELz distributions have lower energies (upper half in each panel), while at $\rm z=0$ clumps at higher $|E|$. The density maps are transparent to distinguish the overlap for some of the most recent mergers. Both the energy and angular momentum of accreted systems are not conserved after the time of accretion, as often assumed in the literature. This energy change is caused by the mass growth of the host galaxy and the angular momentum transformations are due to the non-axisymmetric time-dependent potential of the galaxy with a certain contribution from the population of satellites orbiting the host.}
\label{fig2::mergers_e_lz_evolution}
\end{center}
\end{figure*}
%%%%%%%%%%%%%%%%%%%%%%%%%%%%%%%%%%%%%%%%%%%%%%%%%%%%%%%%%

%%%%%%%%%%%%%%%%%%%%%%%%%%%%%%%%%%%%%%%%%%%%%%%%%%%%%%%%%
\begin{figure*}[t!]
\begin{center}
\includegraphics[width=0.8\hsize]{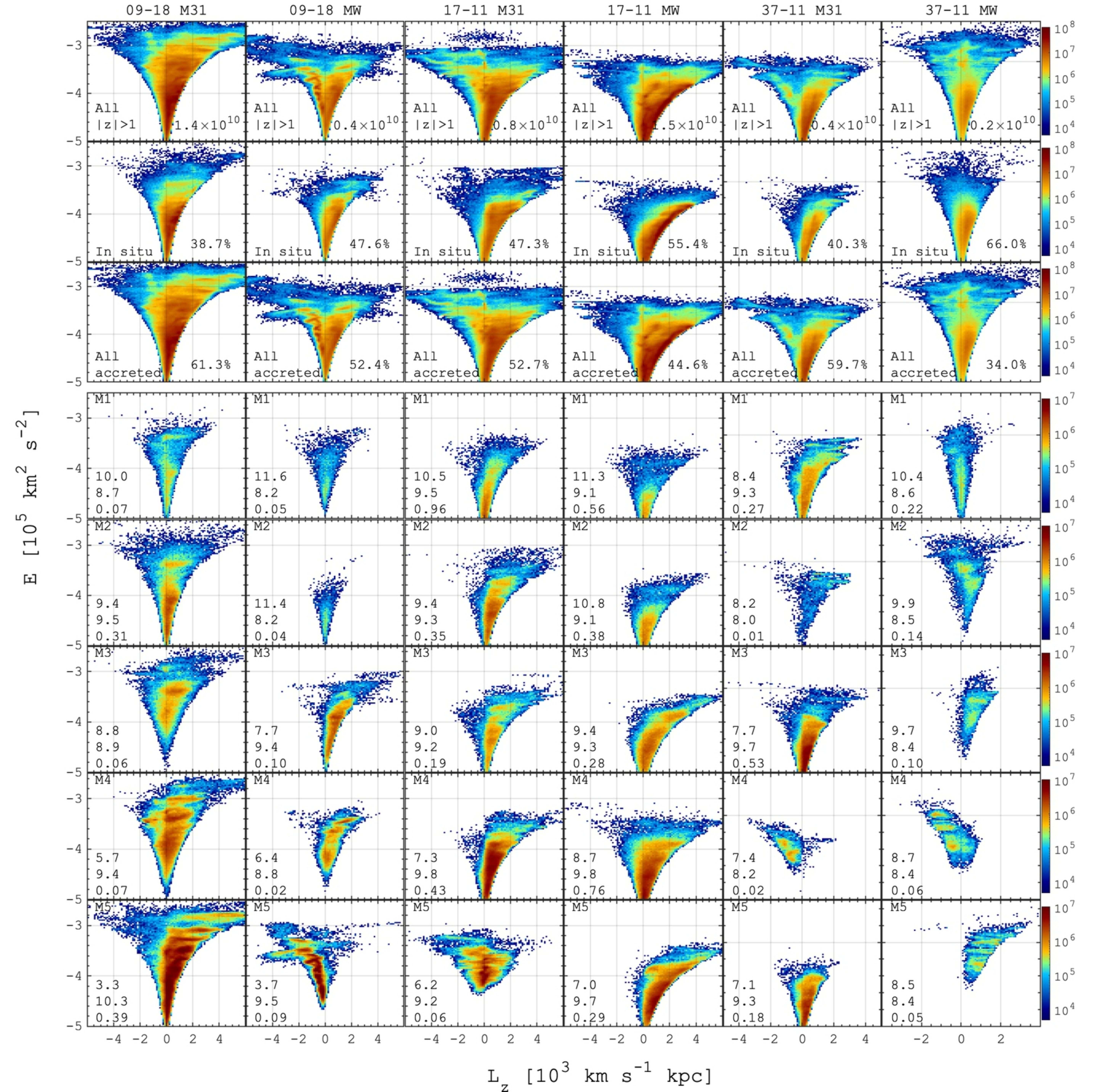}
\caption{Energy-angular momentum space. {\it Top three rows:} Energy-angular momentum relation for all stars~(first row), In situ stars~(main progenitor, second row), and accreted stars~(third row) at $\rm z=0$. To focus on the stellar halo, only stars with $|z|>1$~kpc (away from the disc  mid-plane) were selected. The corresponding stellar mass  is shown in the bottom right of each panel. {\it Bottom five rows:} \ELz relation for the five most significant mergers at $\rm z=0$~(M1,...,M5; see Fig.~4 in \citetalias{KhoperskovHESTIA-1}). The merger accretion lookback time~(Gyr), total stellar mass of the merger debris at the time of the merger ($\rm log_{10}(M_{*}/M_\odot)$), and the stellar mass ratio  ($\mu_{*}$) relative to the main M31/MW progenitor at the time of the merger are given in the bottom left corner of each panel with individual mergers debris. All colour bars represent stellar density, in units of $\rm \Msun$. The in situ stars, being kicked out from the plane, show some net rotation with some overdensities and features similar to those of the accreted stars. The latter contain a number of overdensities, which are the relics of the different merger events. Some individual merger remnants are represented by a number of features and overdensities.}\label{fig2::e_lz_all}
\end{center}
\end{figure*}
%%%%%%%%%%%%%%%%%%%%%%%%%%%%%%%%%%%%%%%%%%%%%%%%%%%%%%%%%

%%%%%%%%%%%%%%%%%%%%%%%%%%%%%%%%%%%%%%%%%%%%%%%%%%%%%%%%%
\begin{figure*}[t!]
\begin{center}
\includegraphics[width=1\hsize]{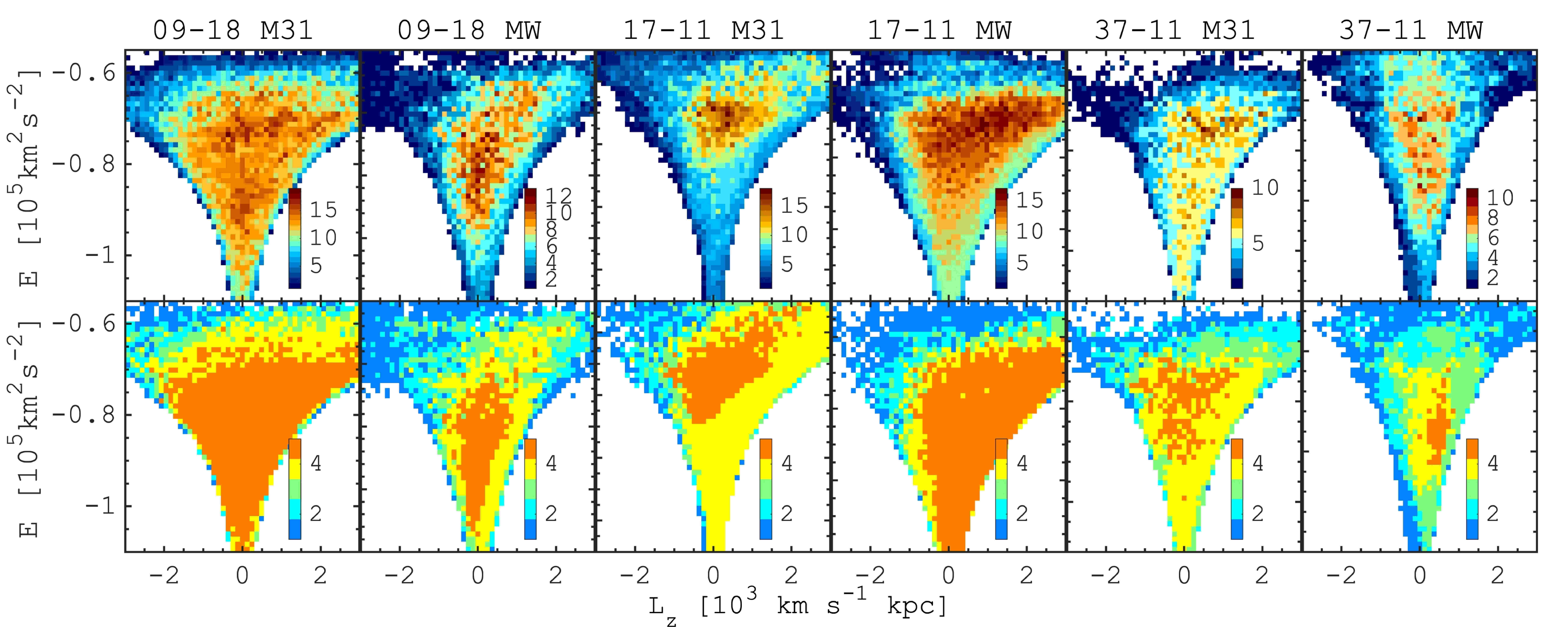}
\caption{Overlap of debris from different mergers in the \ELz plane. Each pixel is coloured by the number of merger events contributing to that region, as given in the colour bar. {\it Top row:} Debris from all mergers for each HESTIA galaxy, as indicated above each panel. The largest overlap of mergers are shown in maroon. {\it Bottom row:} Same as top row, but limited to the five most significant mergers. These maps suggest that the most crowded region of the \Elz space corresponds to the non-rotating region, which is where the Gaia-Enceladus merger was identified~\citep{2018Natur.563...85H}. The number of merger remnants decreases in the regions of substantial rotation.}
\label{fig2::number_of_remnants}
\end{center}
\end{figure*}
%%%%%%%%%%%%%%%%%%%%%%%%%%%%%%%%%%%%%%%%%%%%%%%%%%%%%%%%%

%%%%%%%%%%%%%%%%%%%%%%%%%%%%%%%%%%%%%%%%%%%%%%%%%%%%%%%%%%
\begin{figure*}[t!]
\begin{center}
 \includegraphics[width=1\hsize]{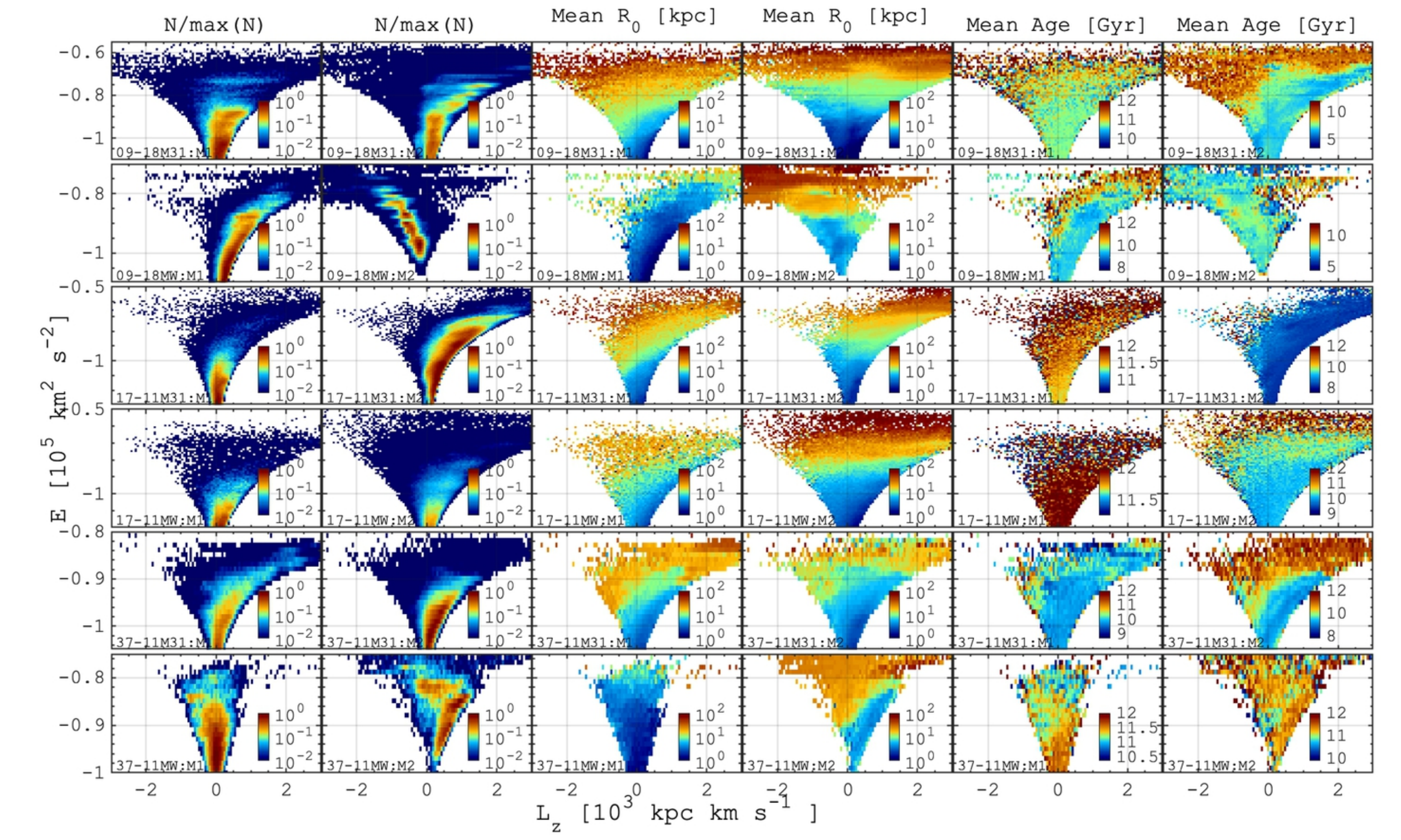}
 \caption{Energy-angular momentum relation for each of the two most significant mergers at $\rm z=0$ per galaxy in our sample. The two left columns show the density distributions of each of the two mergers, the two middle columns are colour-coded by the mean distance ($R_0$) of stars inside the dwarf galaxies before they accreted, and the two right columns show the mean stellar age. This figure demonstrates that even a single merger event could result in a number of overdensities in the $E-L_z$ space. More recent events tend to show more substructures; however, even the merger remnants from $9-10$~Gyr ago are not featureless at $\rm z=0$. The $R_0$-gradient suggests how the internal structure of dwarf galaxies is being mapped into the \ELz space once the system is accreted.  }\label{fig2::ind_mergers_e_lz_ages_dist}
 \end{center}
 \end{figure*}
%%%%%%%%%%%%%%%%%%%%%%%%%%%%%%%%%%%%%%%%%%%%%%%%%%%%%%%%%

Analysis of the MW halo stars orbits suggest that GSE arrived in the Galaxy on a highly retrograde orbit~\citep{2018Natur.563...85H,2021arXiv210303251N}; however, recently \cite{2021arXiv210800010V} highlighted the importance of radialization of the stellar orbits due to angular momentum exchange between the host and massive satellite, thus making it difficult to constrain the actual orbit of the GSE. This result also suggests that the individual merger debris could lead to the appearance of a number of different phase-space (or energy-angular momentum) features\citep[see, e.g.][]{2000MNRAS.319..657H,2007MNRAS.381..987C, 2009ApJ...694..130M, 2010MNRAS.408..935G,2013MNRAS.436.3602G,2017A&A...604A.106J,2019MNRAS.487L..72G,2021arXiv210303251N} because the satellite stars lost at different pericentre passages can escape their progenitor with different mean energies or angular momenta~\citep[see also the case of leading/trailing tidal debris analysis in cosmological context in][]{2019MNRAS.490L..32S}. \cite{2013MNRAS.436.3602G} showed that the fraction of the halo substructures that can be resolved using the integrals of motions and purely kinematic space varies from $31\%$ to $82\%$ while the rest of accreted mass is smoothly distributed in the phase space. 

Since the revolutionary discovery of the GSE merger, the existence of several other merger remnants are being discussed in the literature, for example Sequoia~\citep{2019MNRAS.488.1235M}, Thamnos~\citep{2019A&A...631L...9K}, Arjuna structure~\citep{2020ApJ...901...48N} and others~\citep[see, ][for review]{2020ARA&A..58..205H,2022ApJ...926..107M}. All the above-mentioned discoveries reveal a number of mergers remnants with similar~(within the uncertainties) chemical abundances and substantial overlap in the phase-space and \ELz coordinates, thus making it difficult to uncover the actual MW assembly history~\citep{2021arXiv210904059B,2021MNRAS.508.1489F,2022arXiv220404233H}. However, some hints on how to understand the complexity of the observational data can be taken from the simulations. A nearly complete MW assembly history has been predicted by using the E-MOSAICS simulations, where artificial neural networks have been used to link the phase-space parameters of the globular clusters in cosmological simulations with those measured in the MW~\citep{2019MNRAS.486.3180K,2020MNRAS.498.2472K}. 

Thanks to the advances in subgrid physics implementation and increased spatial and mass resolution, modern cosmological hydrodynamic simulations make possible to study various galaxy-scale properties of accreted and in situ stellar populations~\citep[see, e.g.][]{2015MNRAS.446..521S, 2016ApJ...827L..23W,2016MNRAS.457.1931S, 2018MNRAS.473.4077P} (see also the recent review by \cite{2020NatRP...2...42V}. Using the EAGLE suite of simulations, \cite{2019MNRAS.482.3426M} studied the chemical composition of accreted satellites and the orbital eccentricity of their stellar remnants, suggesting that the low-$\alpha$ MW halo populations is associated with the debris of massive $\rm 10^{8.5}-10^9\Msun$ satellites, which is quite unusual for the MW-type galaxies in EAGLE. Auriga cosmological simulations of the MW-type galaxies~\citep{2017MNRAS.467..179G} show that a small number of relatively massive destroyed dwarf galaxies dominate the mass of stellar haloes where $90\%$ percent of the mass in the inner $20$~kpc is contributed by only three massive progenitors~\citep{2020MNRAS.497.4459F, 2019MNRAS.485.2589M}; however, the fraction of accreted stars is well below $1\%$ in the galactic center~\citep[see also][]{2020MNRAS.494.5936F}. \cite{2019MNRAS.484.4471F} find that around one-third of the simulated galaxies have a significant population of stars which resemble the Gaia-Sausage structure at low-\FeH. \cite{2020MNRAS.497.1603G} found evidence that the Splash-like component is produced from a gas-rich merger that dynamically heats stars from the MW proto-disc onto halo-like orbits~\citep[see, also,][]{2020MNRAS.494.3880B}. 

Although, the mounting growth of evidence is in favour of a large number of building blocks of the MW stellar halo, their disentangling from each other and from the in situ stars remains uncertain. Therefore, the aim of the paper is to investigate the structure of ancient mergers debris in a set of HESTIA\footnote{https://hestia.aip.de} cosmological simulations of the Local Group~(LG). In particular, we analyse a set of six galaxies from three high-resolution HESTIA simulations~\citep{2020MNRAS.498.2968L}. These simulations were tailored to reproduce the LG galaxies populations, which resemble both realistic M31/MW galaxies in terms of their halo mass, stellar disc mass, morphology separation, relative velocity, rotation curves, bulge-disc morphology, satellite galaxy stellar mass function, satellite radial distribution, and the presence of a Magellanic cloud like objects~\citep{2020MNRAS.498.2968L}, thus making the HESTIA simulations the best tool for studying the assembly history relevant for the M31/MW galaxies. 

In a series of works based on a new set of the HESTIA high-resolution cosmological simulations of the LG galaxies we investigate the impact of the ancient mergers on the in situ stellar populations~\citep[][hereafter \citetalias{KhoperskovHESTIA-1}]{KhoperskovHESTIA-1} and the chemical abundance patterns as a function of stellar ages and kinematics of both accreted and in situ stellar populations~\citep[][hereafter \citetalias{KhoperskovHESTIA-3}]{KhoperskovHESTIA-3}. In this paper, we investigate the present-day phase-space structure and evolution over time of the mergers debris in six M31/MW analogues from the high-resolution hydrodynamical HESTIA simulations. The paper is structured as follows. The HESTIA simulations are briefly described in Section~\ref{sec2::all_model}, where we also provide the merger histories and morphology of the stellar debris. In Section \ref{sec2::assembly_history} we present stellar debris of the mergers in the integrals of motions. In Section~\ref{sec2::sausage} we present the Gaia-Sausage-like features in the HESTIA galaxies. Section ~\ref{sec2::actions} describes the action space and orbital properties of the mergers debris. In Section~\ref{sec2::dualism} we discuss the dual nature of the kinematically defined stellar haloes. Finally, in Section~\ref{sec2::concl} we summarize our main results.

%%%%%%%%%%%%%%%%%%%%%%%%%%%%%%%%%%%%%%%%%%%%%%%%%%%%%%%%%

\section{Model}\label{sec2::all_model}

\subsection{HESTIA simulations}

In this work we analyse the three highest resolution HESTIA simulations of the LG. Each simulation is tailored to reproduce a number of the LG properties~\citep{2020MNRAS.498.2968L}, including the massive disc  galaxies resembling the MW and Andromeda analogues with the population of smaller satellites at $\rm z=0$. 

The HESTIA simulations are performed by using the AREPO code~\citep{2005MNRAS.364.1105S,2016MNRAS.455.1134P}, where gravitational forces are computed using a hybrid TreePM technique~\citep{2005MNRAS.364.1105S}. The HESTIA simulations use the galaxy formation model from~\cite{2017MNRAS.467..179G}, which is based on the Illustris model~\citep{2013MNRAS.436.3031V}, and implements the most important physical processes relevant for the formation and evolution of galaxies.  HESTIA simulations assume a cosmology consistent with the best fit values~\citep{2014A&A...571A..16P}: $\sigma_8 = 0.83$ and $\rm H0 = 100~h\, km\, s^{-1} Mpc^{-1}$ where $h = 0.677$. We adopt $\Omega_\Lambda = 0.682$ throughout and $\Omega_M = 0.270$ and $\Omega_b = 0.048$. Haloes and subhaloes are identified at each redshift by using the publicly available AHF\footnote{http://www.popia.ft.uam.es/AHF} halo finder~\citep{2009ApJS..182..608K}. For more details we refer the reader to the HESTIA simulations introductory paper~\citep{2020MNRAS.498.2968L} and \citetalias{KhoperskovHESTIA-1}.

In the rest of the paper, the term `in situ' refers to the stars formed in the most massive M31/MW galaxy progenitor, while `accreted' refers to stars formed in other galaxies and then accreted onto the most massive one~\citep[see, e.g.][]{2020MNRAS.497.4459F,2021MNRAS.503.5826A}.

%%%%%%%%%%%%%%%%%%%%%%%%%%%%%%%%%%%%%%%%%%%%%%%%%%%%%%%%%
\begin{figure*}[t!]
\begin{center}
\includegraphics[width=1\hsize]{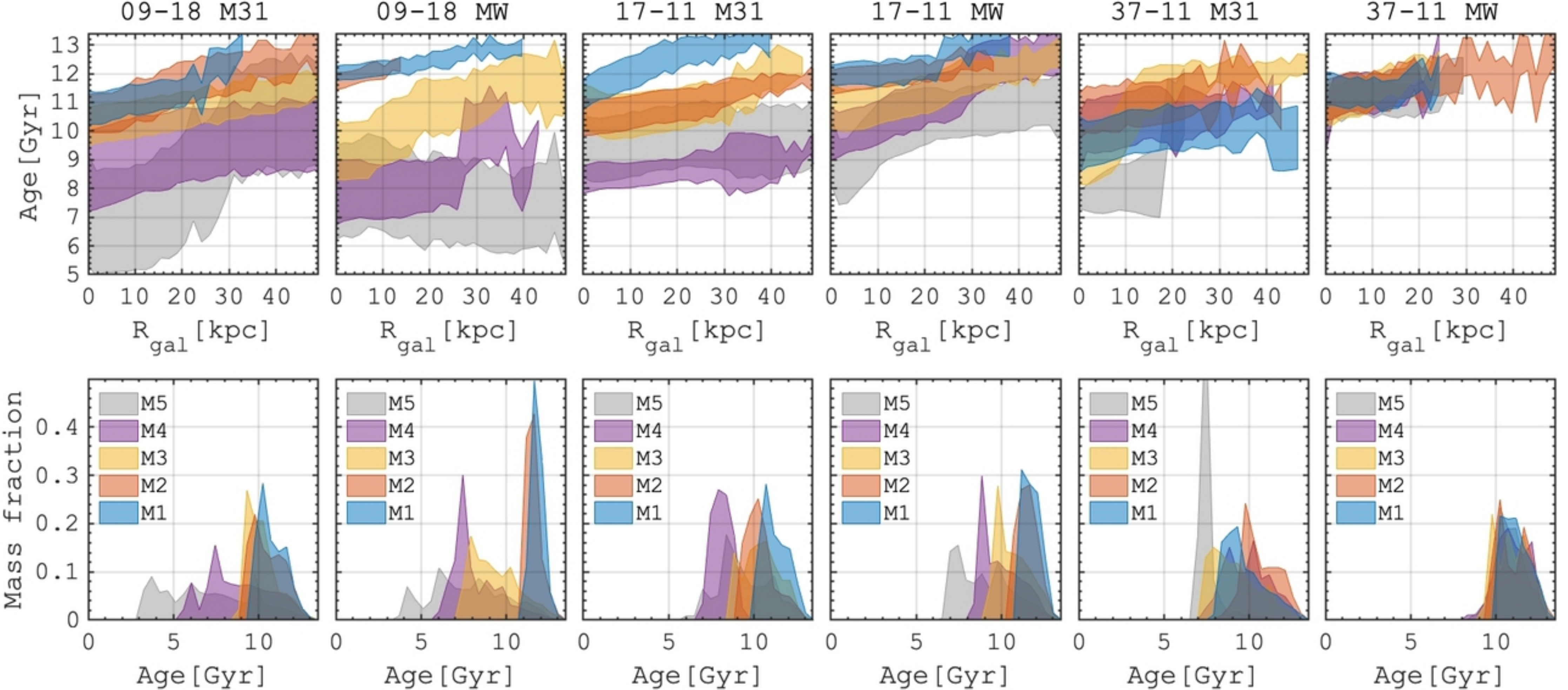}
\caption{Age structure of the accreted halo components. {\it Top:} Mean age of stars accreted from the five most significant mergers as a function of galactocentric distance ($\rm R_{gal}$) at $\rm z=0$. {\it Bottom:} Corresponding age distributions, colour-coded as above. The strongest burst of star formation is typically for the youngest stars in a given merger, thus corresponding to the time of accretion. Age estimates for accreted stars can therefore be used to time each merging event. }\label{fig2::ages_of_mergers}
\end{center}
\end{figure*}
%%%%%%%%%%%%%%%%%%%%%%%%%%%%%%%%%%%%%%%%%%%%%%%%%%%%%%%%%

%%%%%%%%%%%%%%%%%%%%%%%%%%%%%%%%%%%%%%%%%%%%%%%%%%%%%%%%%
\begin{figure*}[t!]
\begin{center}
\includegraphics[width=1\hsize]{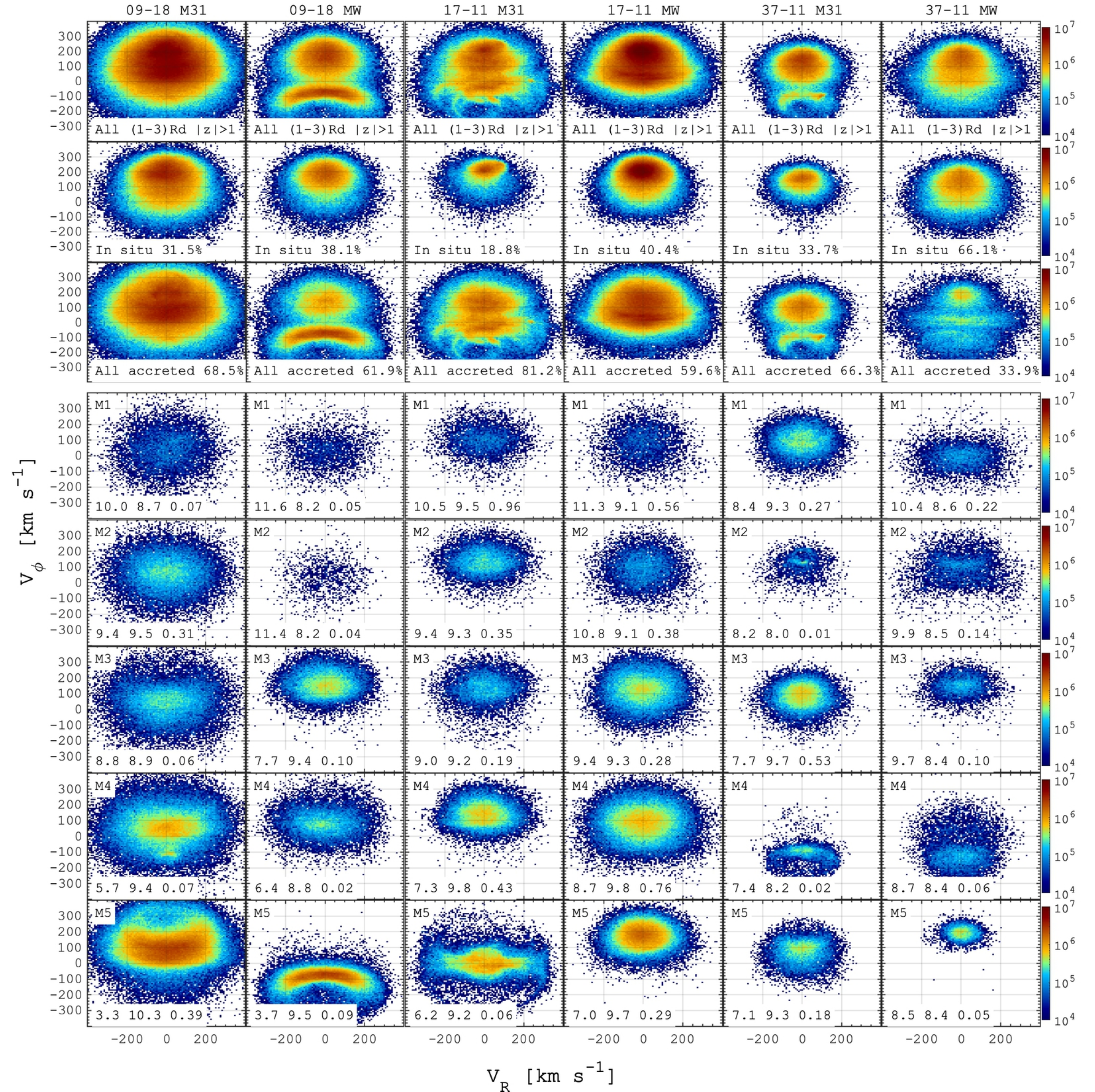}
\caption{Kinematic structure of the stellar haloes. {\it Top three rows:} Stellar density distributions in the \uv plane at redshift zero for stars in the range  $(1-3) R_d$~(where $R_d$ is the disc scale-length  from \cite{2020MNRAS.498.2968L}) and $|z|>1$~kpc including all stars (first row), only in situ stars~(second row), and only accreted populations~(third row). The mass fractions of the in situ and accreted populations are shown at the bottom of the second and third rows, respectively. {\it Bottom five rows:} As above, but for the five most significant mergers at $\rm z=0$~(M1-M5; see Fig.~4 in \citetalias{KhoperskovHESTIA-1}). The merger accretion lookback time~(Gyr), the total stellar mass of the merger debris at the time of the merger ($\rm log_{10}(M_{*}/M_\odot)$), and the stellar mass ratio ($\mu_{*}$) relative to the main M31/MW progenitor at the time of the merger are given at the bottom of each panel. The colour bar represents a normalized density relative to the maximum value. All M31/MW HESTIA galaxies demonstrate the presence of Gaia-Sausage-like features, which exhibit different individual shapes and radial velocity ranges.}\label{fig2::uv}
\end{center}
\end{figure*}
%%%%%%%%%%%%%%%%%%%%%%%%%%%%%%%%%%%%%%%%%%%%%%%%%%%%%%%%%

%%%%%%%%%%%%%%%%%%%%%%%%%%%%%%%%%%%%%%%%%%%%%%%%%%%%%%%%%
\begin{figure*}[t!]
\begin{center}
\includegraphics[width=0.95\hsize]{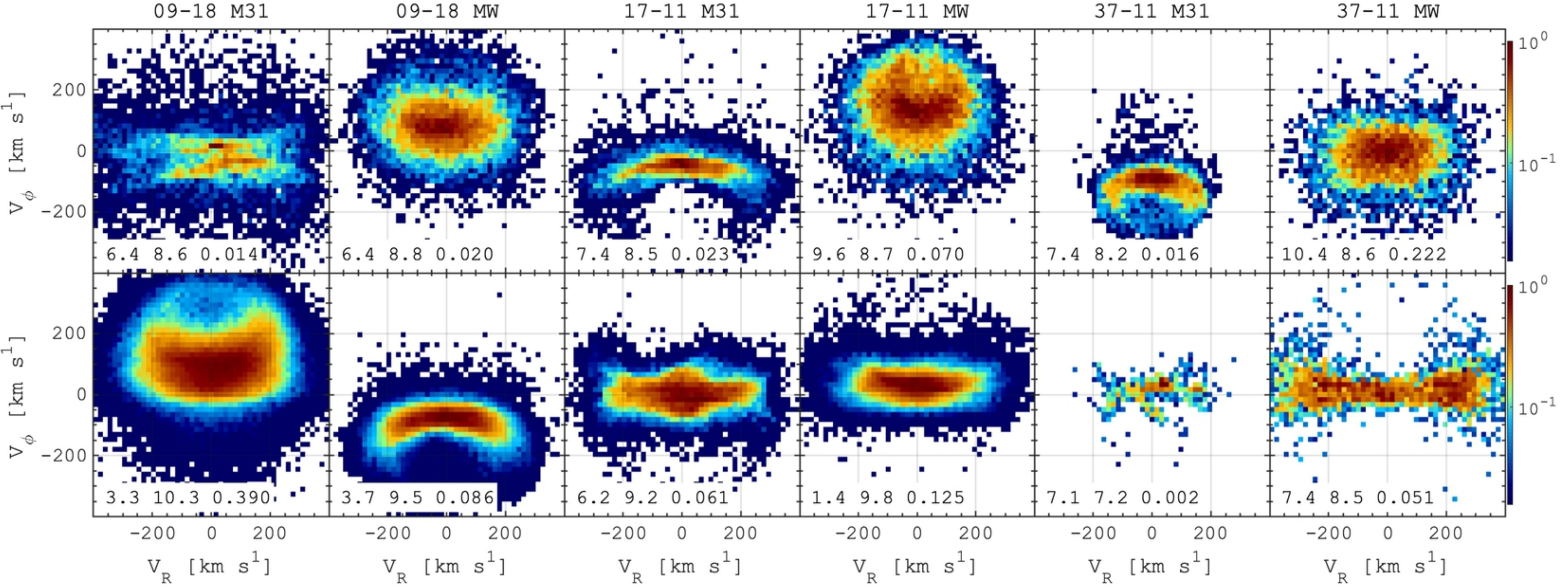}
\caption{Examples of the Gaia-Sausage-like features in the \UV plane for all M31/MW HESTIA galaxies. Each panel corresponds to different merger remnants in the range $|z|>1$~kpc and $(1-3) R_d$ where $R_d$ is the disc scale-length from \cite{2020MNRAS.498.2968L}. The merger accretion lookback time~(Gyr), the total stellar mass of the merger debris at the time of the merger ($\rm log_{10}(M_{*}/M_\odot)$), and the stellar mass ratio ($\mu_{*}$) relative to the main M31/MW progenitor in the time of the merger are given at the bottom of each panel.}\label{fig2::sausage_evolution}
\end{center}
\end{figure*}
%%%%%%%%%%%%%%%%%%%%%%%%%%%%%%%%%%%%%%%%%%%%%%%%%%%%%%%%%

%%%%%%%%%%%%%%%%%%%%%%%%%%%%%%%%%%%%%%%%%%%%%%%%%%%%%%%%%
\begin{figure*}[t!]
\begin{center}
\includegraphics[width=1\hsize]{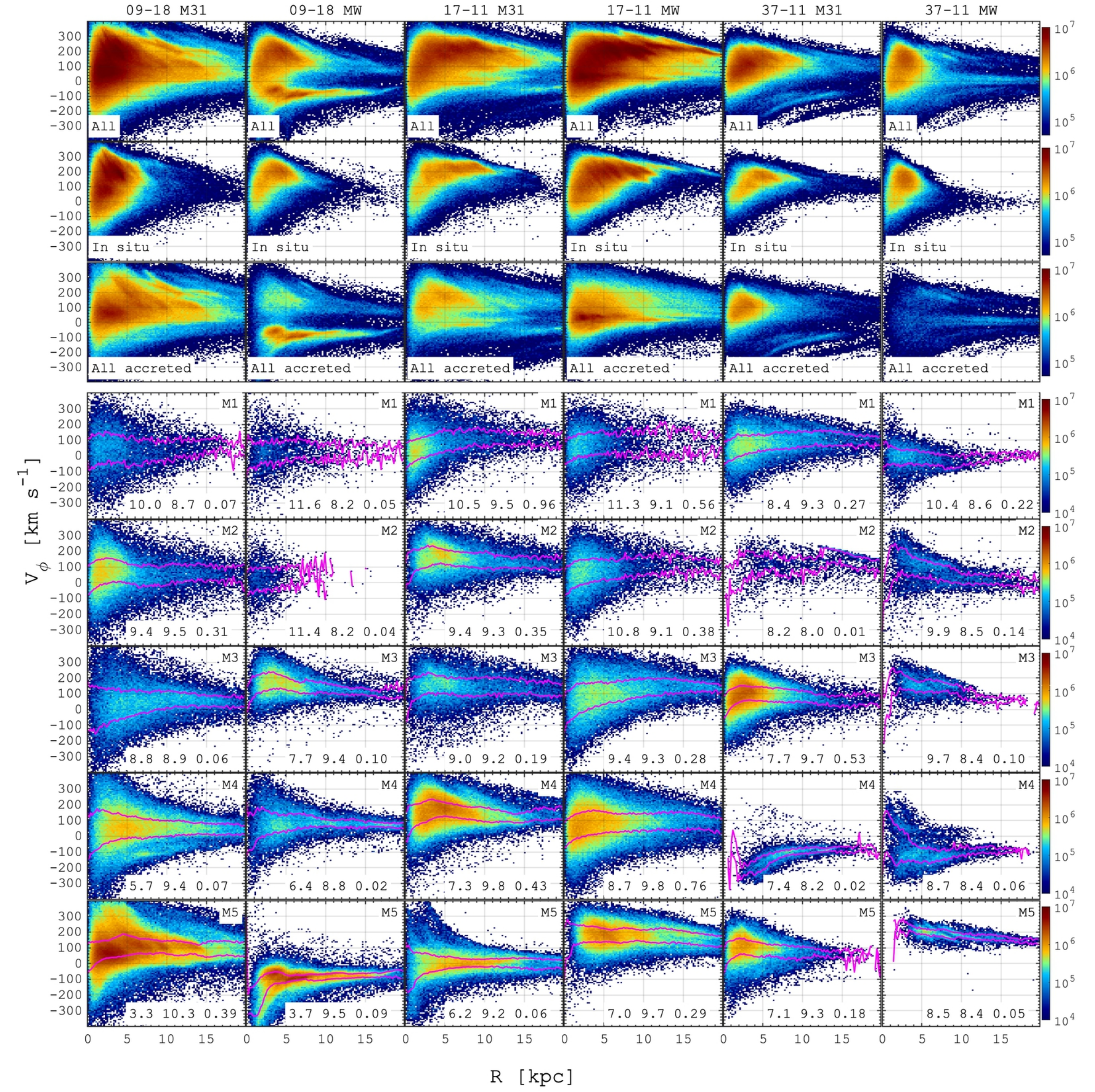}
\caption{Radial structure of the azimuthal velocity distribution. Different stellar populations are shown: all stars~(first row), in situ~(second) and accreted populations~(third row) taken $>1$~kpc away from the galactic plane. All other rows below~(fourth to eighth) correspond to the most significant merger remnants. The merger accretion lookback time~(Gyr), the total stellar mass of the merger debris at the time of the merger ($\rm log_{10}(M_{*}/M_\odot)$), and the stellar mass ratio ($\mu_{*}$) relative to the main M31/MW progenitor at the time of the merger are given at the bottom of each panel. Magenta lines in the individual merger debris distribution show the dispersion level along the radius. The in situ stars rotate faster around the galactic centre and show the presence of diagonal ridges similar to those discovered in the MW~\citep{2018Natur.561..360A}, which are likely to be the manifestation of spiral arms~\citep{2021arXiv211115211K} and/or bar resonances~\citep{2019MNRAS.488.3324F}.  The accreted stars depict a number of radially extended structures with no net rotation and multiple co-rotating and counter-rotating components and small-scale overdensities. The important feature of the individual merger debris is that the azimuthal velocity dispersion remains nearly constant along the galactocentric distance, except for the innermost parts of the galaxies.}\label{fig2::r_vphi}
\end{center}
\end{figure*}
%%%%%%%%%%%%%%%%%%%%%%%%%%%%%%%%%%%%%%%%%%%%%%%%%%%%%%%%%

%%%%%%%%%%%%%%%%%%%%%%%%%%%%%%%%%%%%%%%%%%%%%%%%%%%%%%%%%
\begin{figure*}[t!]
\begin{center}
\includegraphics[width=1\hsize]{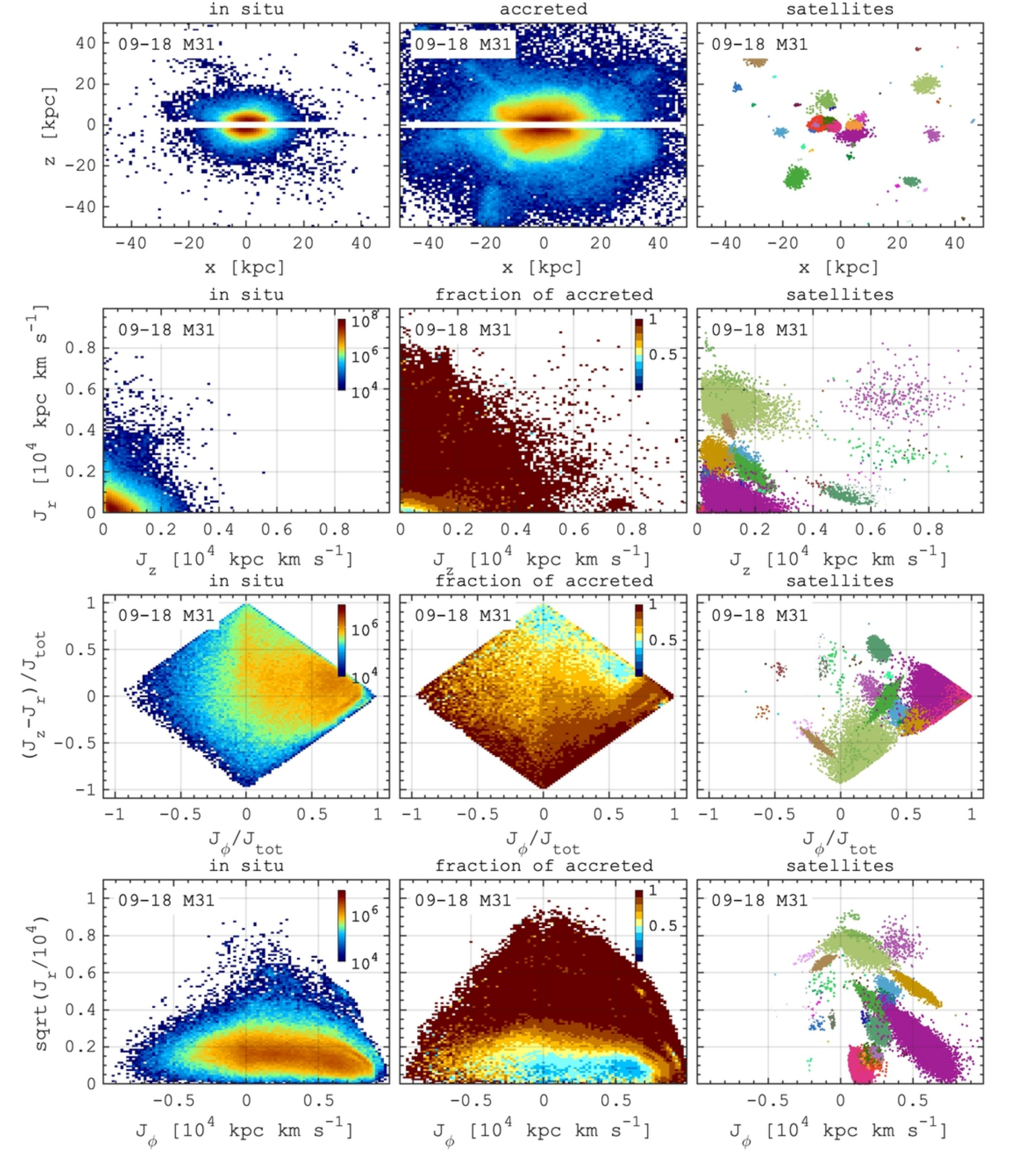}
\caption{Action space analysis. {\it Top:} X-Z projection of in situ stars~(left),  all accreted stars~(middle), and surviving satellites~(right) in 09-18 M31 galaxy at $\rm z=0$. The density maps and the action space analysis are based on the stars~(accreted and in situ) $>1$~kpc from the galactic plane. Different survived satellites are shown in different colours. {\it Second row:} $\rm J_z-J_r$ plane for in situ~(left), the fraction of accreted stellar mass~(centre), and surviving satellites~(right). {\it Third row:} \JJtot plane for in situ~(left), faction of accreted stellar mass~(centre), and survived satellites~(right). {\it Bottom row:} \sJrJp plane for in situ~(left), faction of accreted stellar mass~(center), and survived satellites~(right). While in situ and accreted stars do not show any prominent features in the action space, the dwarf satellites, as expected,  are seen as coherent structures. Accreted stars have on average larger radial and vertical actions, but still substantially overlap with the in situ populations at smaller values. }\label{fig2::actions1}
\end{center}
\end{figure*}
%%%%%%%%%%%%%%%%%%%%%%%%%%%%%%%%%%%%%%%%%%%%%%%%%%%%%%%%%

%%%%%%%%%%%%%%%%%%%%%%%%%%%%%%%%%%%%%%%%%%%%%%%%%%%%%%%%%
\begin{figure*}[t!]
\begin{center}
\includegraphics[width=1\hsize]{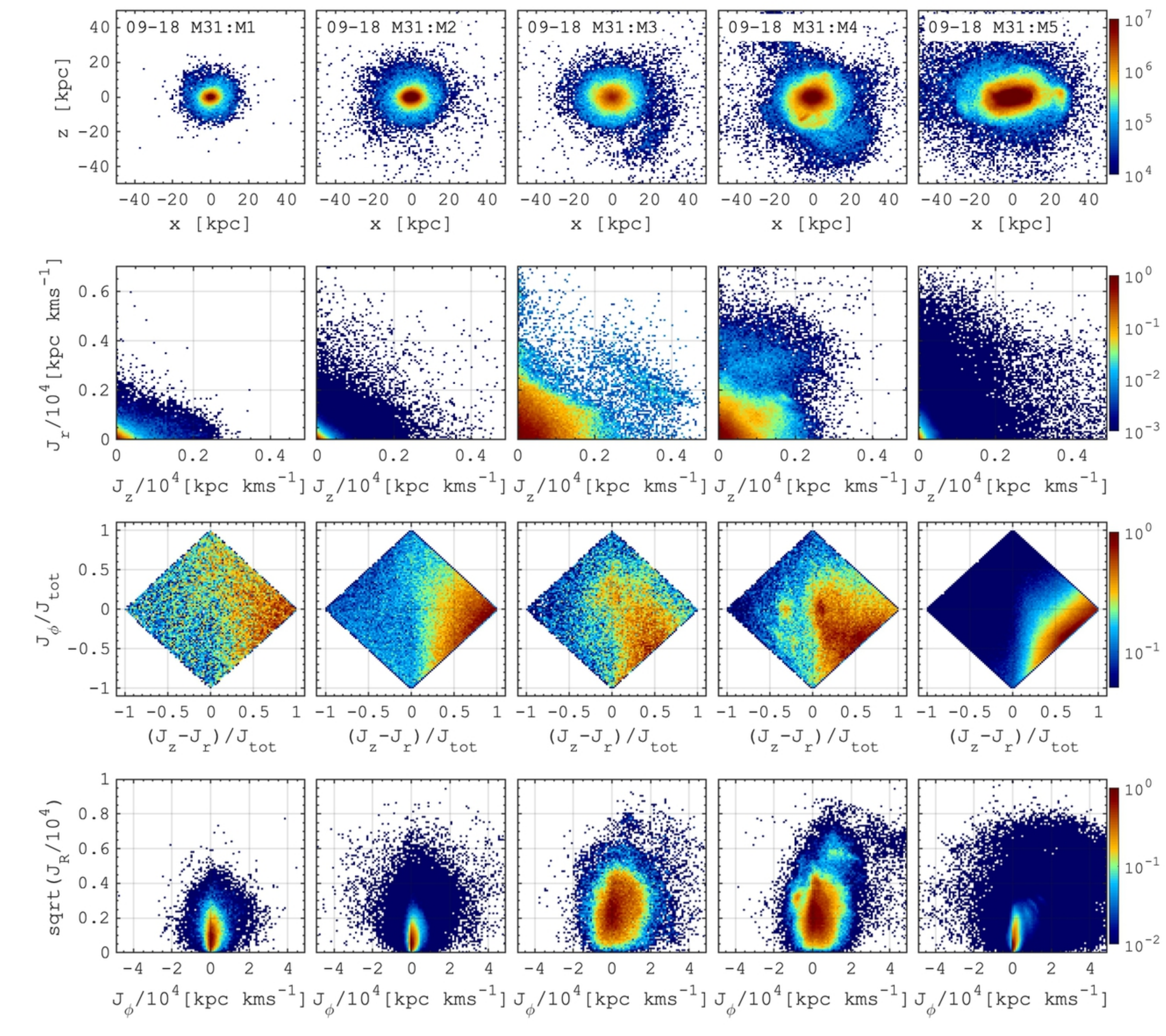}
\caption{Action space for five of the most significant merger remnants in the 09-18 M31 galaxy. From top to bottom: X-Z density maps, $J_z-J_r$ space, \JJtot plane, and \sJrJp plane. Similar to Fig.~\ref{fig2::actions1}~(for the entire accreted population), the individual merger debris completely fill the \JJtot plane; nevertheless, different remnants peak at slightly different regions of the diagram depending on the orbital parameters. The interesting feature, seen in the second row, is that the merger remnants show a different ratio of $J_z$ to $J_r$. Some substructures, however, can be seen at the very large radial and vertical actions, which correspond to the outer parts of the debris where the dynamical time is longer and, thus phase mixing is not yet completed for the most recent mergers at $\rm z=0$.}\label{fig2::actions2}
\end{center}
\end{figure*}
%%%%%%%%%%%%%%%%%%%%%%%%%%%%%%%%%%%%%%%%%%%%%%%%%%%%%%%%%

%%%%%%%%%%%%%%%%%%%%%%%%%%%%%%%%%%%%%%%%%%%%%%%%%%%%%%%%%
\begin{figure*}[t!]
\begin{center}
\includegraphics[width=1\hsize]{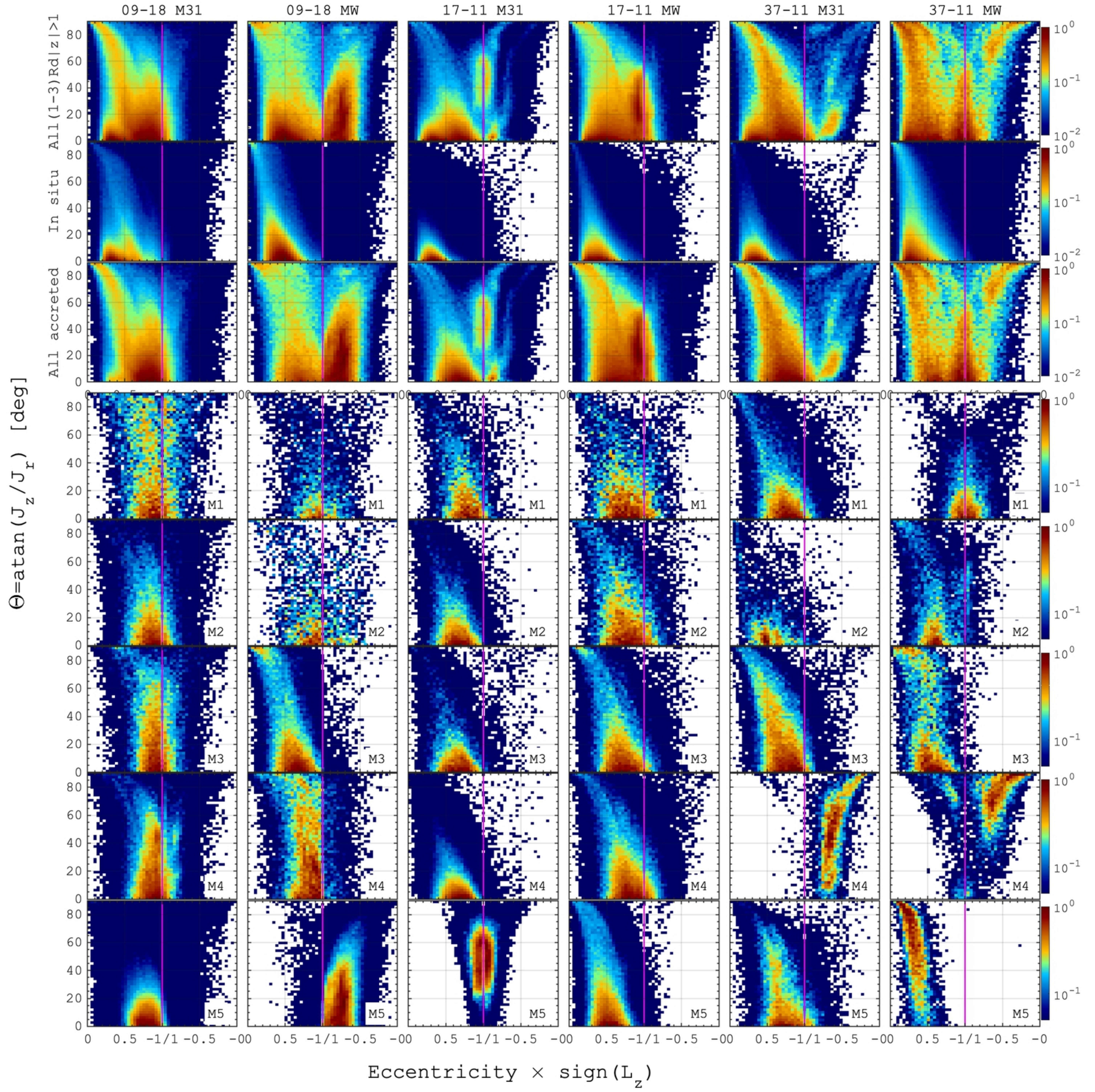}
\caption{Density distribution in the $\Theta = \arctan(J_z/J_r)$-eccentricity plane for all stars~(first row), in situ stars~(second row), and all accreted populations~(third row). All other rows below~(fourth to eighth) correspond to the most significant merger remnants. The eccentricity values are multiplied by the sign of the angular momentum; thus, the counter-rotating stars have negative values of eccentricity and can be found to the right of the vertical magenta lines. To sharpen the structures, only the stars located in the $(1-2)R_d$ radial range, where $R_d$ is the disc scale-length from \cite{2020MNRAS.498.2968L}, are presented here. The figure shows how the anisotropy of actions correlates with the orbital eccentricity for different populations. The in situ stars distribution shows the presence of vertically cold~(low $\Theta$ angles) low-eccentricity stars representing a disc  component; at higher eccentricities, $\Theta$ remains roughly the same. The accreted stars distribution is even more interesting, where there are many substructures, and the overall distribution is different from the in situ stars; moreover, different merger debris, in many cases, show different behaviour. The individual merger debris~(fourth to eighth) do not represent the entire complexity of the $\arctan(J_z/J_r)$-eccentricity space, which is more complex in the third row where all the merger debris are taken into account.}\label{fig2::actions_ecc}
\end{center}
\end{figure*}
%%%%%%%%%%%%%%%%%%%%%%%%%%%%%%%%%%%%%%%%%%%%%%%%%%%%%%%%%

%%%%%%%%%%%%%%%%%%%%%%%%%%%%%%%%%%%%%%%%%%%%%%%%%%%%%%%%%
\begin{figure*}[t!]
\begin{center}
\includegraphics[width=1\hsize]{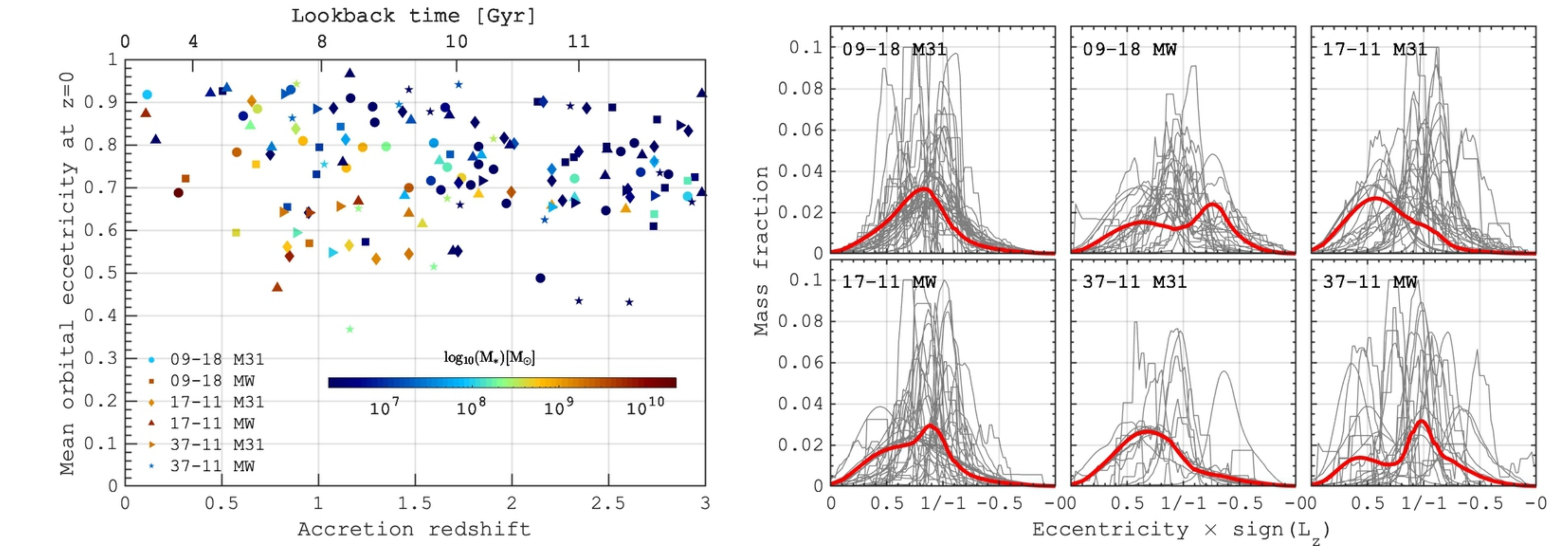}
\caption{Orbital structure of accreted populations. {\it Left:} Mean eccentricity of merger remnants stars as a function of accretion time for all the mergers in the M31/MW HESTIA galaxies. The different symbols correspond to different simulations and galaxies and they are colour-coded by the stellar mass at the time of the merger. {\it Right:} Stellar-mass weighted distributions of the eccentricities for individual merger remnants~(grey) and the mean distribution~(red) at $\rm z=0$. Here, the eccentricity is multiplied by the sign of the angular momentum. Although the stars in the merger debris show a wide range of eccentricities, the mean eccentricity of the debris tends to decrease with time, whereas more recently accreted debris have slightly lower eccentricity. This evidently correlates with the stellar mass of the accreted system, which tends to be higher for the galaxies that accreted earlier, and thus evolved longer.}\label{fig2::eccentricity}
\end{center}
\end{figure*}
%%%%%%%%%%%%%%%%%%%%%%%%%%%%%%%%%%%%%%%%%%%%%%%%%%%%%%%%%

%%%%%%%%%%%%%%%%%%%%%%%%%%%%%%%%%%%%%%%%%%%%%%%%%%%%%%%%%
\begin{figure*}[t!]
\begin{center}
\includegraphics[width=1\hsize]{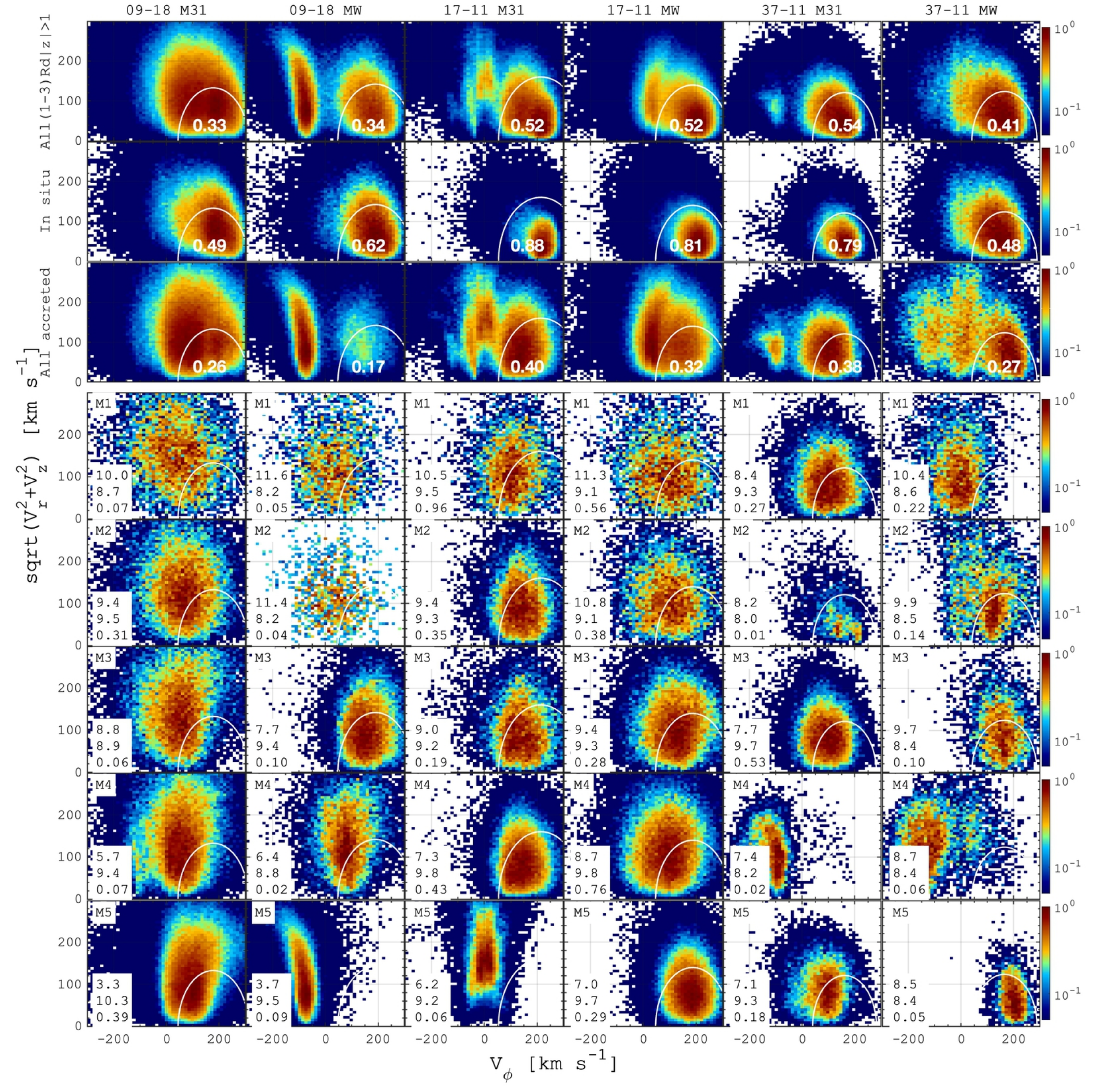}
\caption{Toomre  diagram for all stars in $(1-3)R_d$~($R_d$ is the disc scale-length) and $|z|>2$~kpc~(first row), in situ stars~(second row), and all accreted populations~(third row). The mass fraction of in situ and accreted populations are shown at the bottom in the second and third rows, respectively. {\it The five rows below the gap} correspond to the Toomre diagram for five of the most significant mergers at $\rm z=0$~(M1-M5; see Fig.~4 in \citetalias{KhoperskovHESTIA-1}). The merger accretion lookback time~(Gyr), total stellar mass of the merger debris at the time of the merger ($\rm log_{10}(M_{*}/M_\odot)$), and the stellar mass ratio  ($\mu_{*}$) relative to the main M31/MW progenitor at the time of the merger are given in the bottom left corner of each panel. The colour bar is in $\rm M_{*}$ scale. The white circle corresponds to the thick disc  selection: $V_{circ}180(\kmps)/240(\kmps)$, where $V_{circ}$ is the circular velocity value at $2R_d$ for each galaxy; $180$~\kmps corresponds to the selection made for the MW; and $240$~\kmps is the MW circular velocity. We suggest that the disc-like regions in the Toomre diagram are populated by the accreted stars. A single debris spans over a large volume, and, in some cases, the bulk of the accreted stars have (thick) disc-like kinematics. 
}\label{fig2::toomre_diagram}
\end{center}
\end{figure*}
%%%%%%%%%%%%%%%%%%%%%%%%%%%%%%%%%%%%%%%%%%%%%%%%%%%%%%%%%

%%%%%%%%%%%%%%%%%%%%%%%%%%%%%%%%%%%%%%%%%%%%%%%%%%%%%%%%%
\begin{figure}[t!]
\begin{center}
\includegraphics[width=1\hsize]{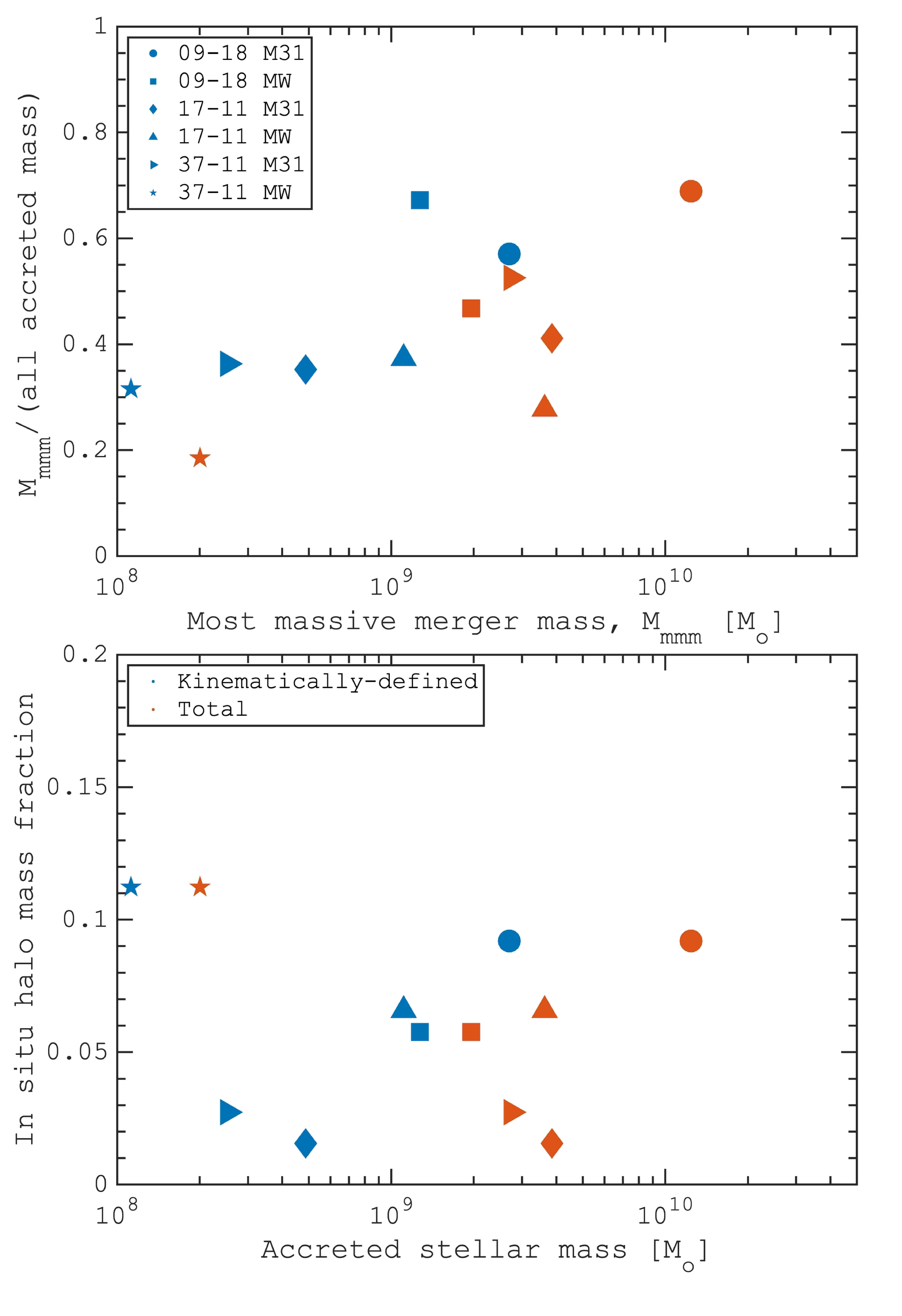}
\caption{Relations between stellar haloes masses and masses of accreted systems. {\it Top:} mass of the most massive merger~($\rm M_{mmm}$) relative to the accreted mass vs mass of the most massive merger. {\it Bottom:} In situ halo mass fraction versus the total accreted stellar mass. Different symbols show different M31/MW HESTIA galaxies. Shown are the kinematically defined components using Toomre diagrams~(in blue), while the total mass without any kinematics selections~(in red). Depending on the definition, the single most massive merger contributes from $20\%$ to $70\%$ of the total accreted mass~\citep[see also, ][]{2010MNRAS.406..744C,2019MNRAS.485.2589M}. At the same, the in situ mass fraction correlates well with the total accreted mass.  }\label{fig2::masses}
\end{center}
\end{figure}
%%%%%%%%%%%%%%%%%%%%%%%%%%%%%%%%%%%%%%%%%%%%%%%%%%%%%%%%%

\subsection{Orbital parameters and actions calculation}\label{sec2::actions_calculation}

For each simulation snapshot we define a coordinate system $(x,y,z)$ centred on the $10\%$ of the most bound in situ star particles and aligned with their principal axes, such that the disc plane of the host galaxy is aligned with the $x-y$ plane. We use galactocentric cylindrical coordinates with velocities $V_{\phi}$, $V_r$, and $V_z$, corresponding to the tangential, radial, and vertical directions. We also use the integrals of motion, focusing on angular momentum in the $z$ direction $L_z$, the total orbital energy per unit mass $E$, and the axisymmetric actions $J_r$, $J_\phi$, and $J_z$. 

To characterize the stellar orbital parameters (eccentricity, apocentre $R_{max}$, and maximum vertical excursion from the disc mid-plane $Z_{max}$), we integrated the orbits of star particles in a smooth potential rather than tracing them across different simulation snapshots. This allowed us to obtain their instantaneous values for each snapshot independently, which is not possible from the direct output data. To calculate the orbital parameters, we first used AGAMA~\citep{2019MNRAS.482.1525V} to compute a smooth version of the gravitational potential. In order to avoid the perturbation of orbits from massive satellites, we interpolated the galaxy potential using the host galaxy particles. The potential due to dark matter and halo gas is represented by a symmetric expansion in spherical harmonics up to $l = 4$, and the potential of the stars and the gaseous disc is approximated by an azimuthal harmonic expansion up to $m = 4$. We next integrated the orbits of star particles in this potential for $20$ Gyr. This long timescale was chosen to account for halo particles with small orbital frequencies. The orbits were integrated using an eighth-order Runge-Kutta DOP853 integrator with an adaptive time step from AGAMA~\citep{2019MNRAS.482.1525V}.

Once the approximation for the gravitational potential is computed, we use the positions and velocities of star particles to calculate three actions $J_r$, $J_\phi$~(same as angular momentum $L_z$), and $J_z$. The three actions are evaluated approximately using the Stäckel fudge method~\citep{2012MNRAS.426.1324B} from AGAMA, which, thus delivers only axisymmetric actions. However, \cite{2021MNRAS.500.2645T} demonstrated that axisymmetrically estimated actions is a powerful diagnostic tool even for the analysis of stellar orbits in a disc of a barred galaxy. Therefore, we suggest that for the purpose of the comparison of accreted and in situ halo stellar populations, the adopted approximation is reasonable~\citep[see, also][]{2021arXiv210408185W}.

\section{Merger remnants in the energy-angular momentum space}\label{sec2::assembly_history}
\subsection{Morphology of the merger debris}
Before analysing the merger debris in various phase-space and energy-momentum coordinates, we take a look at the morphology of the merger remnants at the time of the accretion and at redshift zero. In Fig.~\ref{fig2::mergers_initial} we show the $\rm X-Z$  stellar density maps~(in situ stellar disc is placed in the $\rm X-Y$ plane) of five of the most significant mergers~(M1-M5; see Fig.~4 in \citetalias{KhoperskovHESTIA-1} for the mass-time of accretion) for each M31/MW galaxy at the time of accretion~(i.e. when the satellite can no longer be identified as a bound structure). In each panel, we mark the lookback time of accretion, stellar mass at the time of accretion $\rm \log_{10}(M_{*})$ and the stellar mass ratio $\mu_{*}$ relative to the main M31/MW progenitor. In Fig.~\ref{fig2::mergers_initial} the earlier mergers are shown at the top. There is a diverse morphology of the merger structures; in particular, we can see several examples of tidal tails and shell-like structures, a few quasi-circular streams or loops, but most of the mergers are seen as smooth quasi-spherical stellar density distribution with a prominent density peak~(the core of a dwarf galaxy). We note that earlier mergers tend to be smoother at the time of accretion, while more recent ones exhibit streams, tails and in general a more complex morphology. This morphological diversity of the remnants is likely because the galaxies that merged at early epochs simply do not have enough time to evolve into internally complex objects. In addition, to be accreted earlier, they should appear relatively close to the main progenitor with a peculiar velocity that is much lower than the escape velocity. The galaxies that merged later have more time to grow and, thus they can have a more complex and long-term orbital evolution around the host galaxy~(M31 or MW).

At redshift zero most of the merger debris look flattened towards the galactic plane of the host galaxy, and appear as smooth stellar ellipsoids~(see Fig.~\ref{fig2::mergers_final}). Only a few debris are not featureless, which is typical behaviour for unrelaxed remnants from relatively recent mergers. Another interesting behaviour is that more ancient mergers tend to be more compact, while more recent debris extend farther away from the center of the host galaxy~\citep{2020MNRAS.499.4863P}. This picture can be understood if the earlier mergers happened when the main progenitor was less massive, and the gravitational potential well was relatively shallow. Later on the mass of the galaxy increases together with gravitational forces, which squeeze the merger debris over time. Obviously, this effect should lead to the change in the total energy of the debris and its transformation in the energy-angular momentum space.

In the following sections we analyse the structure of the most significant merger debris in the HESTIA galaxies in different phase-space, integrals of motions and action space used to uncover the composition of the stellar halo of the MW.

\subsection{Evolution of the \ELz distribution}
We demonstrated that since the merger time, the morphology of the merger debris changes drastically, where coherent tidal structures and streams are completely phase-mixed~(see Figs.~\ref{fig2::mergers_initial} and ~\ref{fig2::mergers_final}). We recall that in the case of axisymmetric and time-independent gravitational potential, a given merger debris is expected to remain clustered in the \ELz coordinates if the dynamical friction is not taken into account~\citep{2017A&A...604A.106J}. However, we already showed that the main M31/MW progenitors acquire a substantial amount of their mass since the very first mergers~(see Fig.~4 in \citetalias{KhoperskovHESTIA-1}), suggesting that, at least, the energy of the stellar merger remnants can be changed and can thus affect the initial phase-space configuration.

In order to study the impact of the galaxy mass growth on the merger debris in Fig.~\ref{fig2::mergers_e_lz_evolution} we show the \ELz distribution for five of the most significant mergers. Each panel shows the distribution at the time of the merger~(upper distributions, blue contours), and at $\rm z=0$~(bottom distributions, red contours). The vertical displacement in Fig.~\ref{fig2::mergers_e_lz_evolution}, or the energy change, is explained by the mass growth of the host. At early times a dwarf galaxy is accreted into a low-mass M31/MW progenitor and, thus has higher energy in a shallow potential. Later on, the mass of the host galaxy increases (due to accretion of other systems and gas) and the total energy of the debris decreases. Thus, the mass growth of the main progenitor plays an essential role in the evolution of the total energy of any merger remnant.  Another effect contributing to the change in the total energy is dynamical friction which is, however, most important at the early phases of the merger when the dwarf galaxies are still massive and dense enough~\citep{2020MNRAS.491.4591E}.

Another interesting feature seen in Fig.~\ref{fig2::mergers_e_lz_evolution} is that the mergers debris do not conserve their initial~(at the time of accretion) angular momentum $L_z$ distribution. The evolution of the angular momentum distribution can be understood because the galaxies are non-axisymmetric, and the potential is constantly perturbed by orbiting satellites. In this context it is worth mentioning the work by \cite{2022MNRAS.513.1867D} where by using the ARTEMIS set of 45 high-resolution cosmological simulations, the authors found a transformation of the merger debris caused by subsequent accretion of dwarf galaxies. Although most of the mergers show a little net rotation, we can see that the tails of the distribution may change the direction of rotation~(see e.g. M3 mergers in 09-18~M31/MW, M2 and M3 in 17-11~M31 and M4 in 37-11~MW). More generally, most of the different debris contribute to slowly co-rotating or non-rotating stellar components near $L_z \approx 0$ where they overlap with each other. This overlap suggests that different merger remnants cannot be disentangled from each other, at least not by using their distribution in the \ELz plane at $\rm z=0$.

\subsection{Individual mergers and in situ stars in \ELz }

Next, we provide the structure of the \ELz space of the stellar haloes in the M31/MW HESTIA galaxies, including in situ, all the accreted populations and some individual merger debris at $\rm z=0$. In Fig.~\ref{fig2::e_lz_all}~(top) we show the \ELz plane for all the stars $>1$~kpc away from the galactic plane at $\rm z=0$. The in situ stars~(formed inside the main progenitor) are shown in the second row, and the third row corresponds to all accreted populations. Overall, the \ELz distributions look quite similar for all the M31/MW galaxies, which suggests that slightly different accretion and mass growth histories~(see Fig.~4 in \citetalias{KhoperskovHESTIA-1}) result in rather similar halo compositions. Using a simple spatial cut, we find the mass of accreted stellar components in the range of $34-61\%$ of the total stellar mass $>1$~kpc from the galactic plane. Therefore, about a half of the stellar mass is made of the in situ stars~($1.3-8\times10^9$~\Msun).

Interestingly, the \ELz distributions of the accreted stars look very similar to the distributions of all stars,  making an impression that accreted stellar populations visually hide within the in situ halo, which, as we already showed in~\citetalias{KhoperskovHESTIA-1}~\citep[see, also][]{2017MNRAS.472.3722G}, is composed of the disc  stars heated up by the mergers. Compared to the in situ stars, the accreted populations span over the larger range of $L_z$ dominating in the counter-rotating tail~($L_z<0$) of the distribution. Nevertheless, the averaged kinematics of the accreted stars shows slightly co-rotating or non-rotating behaviour, similar to that observed in the MW. A single exception here is the MW galaxy from the 09-18 simulation where a very prominent counter-rotating debris appears~(M5) as the result of a rather recent~($3.7$~Gyr ago) and massive~(stellar mass $10^{9.5}$~\Msun) accretion event.

Another feature is that the accreted stars show a number of small-scale clumps or elongated overdensities~\citep[see, also, ][]{2000MNRAS.319..657H,2007MNRAS.381..987C, 2009ApJ...694..130M, 2010MNRAS.408..935G,2015AJ....150..128R,2022arXiv220412187A} while the distributions of the in situ stars seem to be almost featureless~(except for the very high energies), suggesting that the individual mergers do not produce coherent substructures made of in situ stars~\citep[but see][ regarding the substructures made of in situ stars]{2017A&A...604A.106J}. The individual merger remnants distributions~(rows $4$ to $8$ rows in Fig.~\ref{fig2::mergers_e_lz_evolution}) show that not every particular feature in the \ELz plane corresponds to any particular merger debris. Almost all the merger remnants (with no exceptions) contribute to a non-rotating region~($L_z\approx 0$) and, at the same time, depict a number of different overdensities. Of course, more prominent features are made of the debris from the most recent mergers; however, even the very early mergers (from $9-10$~Gyr ago) show some distinct substructures. In some cases, the individual merger debris depict a sequence of small-scale overdensities or group of clumps along the energy axis while being stretched along the angular momentum.

To discuss in more detail the overlap between the merger debris in Fig.~\ref{fig2::number_of_remnants} we show the \ELz plane colour-coded by a number of merger debris that can be found in a given range of parameters. The top row corresponds to all the merger debris we have identified, while the bottom row shows the maps based on the five most significant M1-M5 mergers. The general trend is that most of the mergers contribute to the parameters of the stellar halo with no prominent rotation, where the GSE merger remnant is found in the MW~\citep[see, e.g.][]{2018MNRAS.478..611B,2018Natur.563...85H,2020MNRAS.494.3880B,2020ApJ...901...48N}, while the edges of the \ELz distribution show the presence of a fewer number of merger debris. Roughly the same picture is observed for the most significant mergers~(see bottom of Fig.~\ref{fig2::number_of_remnants}).

\subsection{Dissecting individual debris in \ELz by age and origin}

Although it is expected that merger debris should be clustered in integrals of motion space~(e.g., \ELz), we have shown that in the HESTIA galaxies, a given merger debris occupies a wide range of parameter space and consists of several prominent overdensities. Next we focus on the variations in parameters across the merger debris in \ELz coordinates. In Fig.~\ref{fig2::ind_mergers_e_lz_ages_dist} for each galaxy in our sample, for simplicity we show two individual merger debris in \ELz coordinates~(first and second rows). The next two columns~(third and fourth) show the distribution of the mean distance relative to the centre of a dwarf galaxy before the merger that results in a given stellar remnant. In other words, these columns show how far from the centre  the stars were inside the dwarf galaxy before it merged into the host galaxy. The last two columns in Fig.~\ref{fig2::ind_mergers_e_lz_ages_dist} show the mean stellar age distribution of the individual mergers in \ELz coordinates. 

Figure~\ref{fig2::ind_mergers_e_lz_ages_dist} highlights that even a single merger can result in several prominent groups of stars isolated from each other in \ELz space. In most cases, these groups can be found along the smooth density distribution stretched along the total energy. The next two columns show that stars in different \ELz parts arrive from different distances inside the dwarf galaxy prior to the merger. In particular, the smallest total energy stars at $\rm z=0$ were captured from the outer parts of dwarf galaxies. The outer parts of the dwarf galaxies being less bound are captured earlier and do not penetrate much to the centre of the host, thus building the (relative) outer parts of the stellar halo. The central parts of the dwarf appear to sink towards the centre of the host galaxy due to dynamical friction where they have the lowest total energy. As a result, these populations dominate in the low-$E$ regions of the \ELz space. 

The mean age maps in Fig.~\ref{fig2::ind_mergers_e_lz_ages_dist}~(rightmost columns) do not imply a unique picture. However, in most cases, we can see the age gradient as a function of the total energy. In particular, relatively older stars from the outer parts of dwarf galaxies tend to occupy the upper parts of the \ELz distribution, while the mean age decreases towards the lower values of the total energy. Coupling this information with the previous result, can explain the observed trends. Before the merger, stars formed earlier in the `upside-down' regime occupy the outer parts of dwarfs galaxies; therefore, when the galaxy is being accreted, these older stars are stripped first and appear at higher energies inside the host. At the same time, the cores of dwarf galaxies could have a more extended star formation history and, thus be younger or, at least, have a larger fraction of younger stars. Therefore, being dragged towards the central parts of the hosts, the stars from the cores of the dwarf galaxies populate lower energy levels where the mean age of stellar debris is relatively smaller. The strength of the age gradient in the \Elz space is rather low, especially for the very early mergers~(see, e.g. the MW galaxy in the 17-11 simulation). However, typical age uncertainties for oldest MW stars may not allow to capture the age gradient of the merger debris, but some parameters of stellar populations can change very quickly with the stellar age, for example chemical abundances~(see the metallicity analysis in \citetalias{KhoperskovHESTIA-3}), and, can thus be detected in the observational data.

To explore a bit more the age variations of the merger debris in Fig.~\ref{fig2::ages_of_mergers}, we show the mean stellar age as a function of the galactocentric distance~(top) and the overall age distributions~(bottom) for the individual merger debris. First, we confirm the results seen in the previous figure where the inner parts of the mergers debris tend to be younger~\citep[see also][]{2019MNRAS.485.2589M}. There are a number of exceptions, probably depending on the details of a particular merger evolution. However, the age gradient tends to be larger for the most recent mergers. Once we consider the overall age distribution~(or the star formation history, see bottom of Fig.~\ref{fig2::ages_of_mergers}), we note a substantial increase in the star formation rate just before the merger. Some merger remnants show a few peaks, possibly related to the pericentric passages~\citep[see, e.g.][]{2021MNRAS.506..531D}. This suggests that, in most cases, the age of the youngest stellar populations can be used as a proxy for the merger time; however, for better estimates the stars from the inner galaxy should be considered~\citep[see also ][]{2019MNRAS.485.2589M,2022MNRAS.513.1867D}.

If the relations we observe in the HESTIA galaxies are relevant to the MW, this may affect some of the conclusions regarding the merger history of the MW. In particular, the mean age of the GSE progenitor can be overestimated because its measurements are based on the stars in the solar suburb~\citep[see, e.g.][]{2019NatAs...3..932G,2021MNRAS.508.1489F,2021arXiv211101669B,2021NatAs...5..640M}; however, the youngest stars of the GSE will likely appear in the innermost galaxy, which is still poorly covered by the existing spectroscopic surveys and the stellar age information is even more limited in that regions. Therefore, once we extend the analysis of the MW stellar halo towards its innermost regions, we may constrain even better the age distribution (and thus the merger time) of the GSE and possibly even infer the star formation history of its progenitor.

\section{Gaia-Sausage-like features in HESTIA galaxies}\label{sec2::sausage}

We already mentioned in the Introduction that the last significant merger~(GSE) in the MW was initially discovered as a horizontally aligned structure in the \uv coordinates with zero net rotation~($V_\phi \approx 0$) for low-\FeH stars~\citep[Gaia-Sausage,][]{2018MNRAS.478..611B}. Therefore, it is worth testing how much the HESTIA simulations are able to reproduce this particular feature in a pure kinematic space. In Fig.~\ref{fig2::uv} we show the density distribution in the \UV-plane for stars located in the radial range of $(1-3)R_d$~(where $R_d$ is the disc  scale length) and $>1$~kpc from the galactic mid-plane. In the figure we also separate in situ~(second row) and accreted stellar populations~(third row). In the adopted region the accreted halo constitutes from $34$ to $81\%$ of the mass, while the rest is represented by the heated in situ stars. First, we find that in the M31/MW HESTIA galaxies in situ stars are represented by a single blob whose tail however spans to $\VV\leq0$~\kmps~(or even below in the case of the M31 galaxy in the 09-18 simulation). The in situ stars distribution is almost featureless and shows a behaviour typical for thick-disc  stars, however, with some asymmetry in the 09-18~M31 and 17-11~M31 galaxies caused by the very recent mergers~(see Fig.~4 in \citetalias{KhoperskovHESTIA-1}).

The accreted stars distributions in the \uv plane is much more interesting. In particular, we note the presence of multiple distinct kinematic components with a certain level of similarity between the galaxies. The morphology here can be described as a blob on top of the horizontally aligned structure(s). The blobs for accreted stars have a wider range of radial velocity range compared to the in situ populations and lower median azimuthal velocities~(except for 37-11 MW where the $\langle\VV\rangle$ are essentially the same). This demonstrates a smooth transition from fast-rotating in situ stars to slower accreted stars. 

The most striking features in the accreted density distribution are the horizontally aligned structure(s). The third row in Fig.~\ref{fig2::uv} shows a diverse morphology of slowly rotating components, which can be described as horizontal features with zero net rotation~(17-11~M31, 17-11~MW, 37-11~MW; similar to the Gaia-Sausage) and negative  net rotation~(37-11~MW); and positively~(09-18~M31) and negatively~(09-18~MW, 37-11~MW) bent moustache-like features. We note that in the 37-11~MW galaxy we can directly identify two prominent features below the high-\VV blob that correspond to different accretion events, while in other galaxies the separation between different mergers is not evident.

In order to decompose the density distribution of the accreted stars in Fig.~\ref{fig2::uv}~(rows 4-8), we show the \uv structure of five of the most significant mergers~(M1-M5) separately. Most of the individual merger remnants can be seen as high-velocity dispersion ovals that have different but mostly positive~(or prograde relative to the host component) bulk rotation. Gaia-Sausage-like horizontally aligned features are rare in a given galaxy, and they tend to correspond to the most recent mergers~(09-18 M31/MW, 37-11~M31). In the 17-11~MW galaxy accreted stars show a Gaia-Sausage-like feature, but the corresponding merger is very recent (1-2~Gyr ago) with the relative mass smaller than M1-M5 mergers. 

The MW galaxy from the 37-11 simulation is an exceptional case because at least four  merger remnants~(M1-M4) are seen as flattened structures in the \uv plane, with different bulk values \VV. In this galaxy the relative mass of the mergers decreases with time, and thus each subsequent merger introduces less perturbation to the structure of the previous one. More likely, the mergers are to be accreted with some angular momentum and settle in the galaxy with a certain bulk \VV. Later on, the following mergers perturb the galaxy by heating up all the previous debris, thus leading to an increase in the velocity dispersion, also in the azimuthal direction. Therefore, the transformation from the horizontally aligned structures in the \uv space can be caused by the subsequent mergers. 

So far, we have presented only the most significant mergers; therefore, the \uv space for accreted stars decomposition is not complete, and the variety of kinematic structures is not fully explored. In order to show a diversity of the accreted populations in the \uv space, in Fig.~\ref{fig2::sausage_evolution} we show two different merger debris in each HESTIA galaxy which result in the structures that are the most similar to the Gaia-Sausage. Although we have a very large sample of mergers~($>60$ for all the galaxies) we can see that only a single merger remnant from the MW in the 37-11~(see bottom right) simulation matches precisely the Gaia-Sausage kinematics.

In order to understand the radial structure of accreted populations and the merger debris in Fig.~\ref{fig2::r_vphi}, we show the stellar density map in \RVphi plane for all the stars, in situ, accreted and individual merger remnants for stars $>1$~kpc away from the disc  plane. The figure reveals a number of radially extended nearly horizontal overdensities, which at large distances from the centre occupy rather narrow \VV ranges, and the azimuthal velocity dispersion~(magenta contours) remains roughly constant along the radius. Some of the debris show a fine structure with some clumps or diagonal overdensities. We note an important difference between these accreted structures and the diagonal \RVphi-ridges, discovered by the \Gaia~\citep{2018Natur.561..360A}. In HESTIA galaxies, accreted components have nearly constant \VV, while centre \RVphi-ridges typically have nearly constant energy or angular momentum~\citep{2018A&A...619A..72R}.  Figure~\ref{fig2::r_vphi} also suggests that the galactic centre is a very crowded region because different mergers debris contribute there. However, the outer disc seems to be a promising region for the analysis of accreted components. Taking into account similarities between some individual merger remnants in Fig.~\ref{fig2::uv} and the Gaia-Sausage structure, we could expect that in the MW, the accreted populations should be seen as radially extended structures with certain values of the azimuthal velocity depending on the orbital parameters of the merger~\citep[see, e.g. recent works by][]{2020ApJ...901...48N,2022arXiv220412187A}.

\section{Action space~($J_r, J_z, J_\phi$) and orbital analysis}\label{sec2::actions}
In this section we analyse the orbital composition and axisymmetric actions of the stellar component of the HESTIA galaxies. For all star particles, we reconstructed their orbital eccentricity by integrating their orbits in the fixed galactic potential at $z=0$ using the AGAMA code~\citep{2019MNRAS.482.1525V} which was also used to compute the actions~($J_r, J_z, J_\phi$; see Section~\ref{sec2::actions_calculation}).

\subsection{Action space for in situ, accreted stars, and satellite galaxies}
In the previous section we demonstrate that the accreted stellar populations have a complex structure in \ELz, \uv, \RVphi. In these kinematic coordinates, the mergers debris overlap with each other and with the in situ stars heated up by merging events. In this section, we focus our analysis on the orbital actions, which are conserved under adiabatic changes of the gravitational potential of a galaxy~\citep{2008gady.book.....B} and, thus, the merger debris should be clustered in the action space. Although we already demonstrated that the last assumption is not exactly applicable for the HESTIA galaxies, it is still worth testing what we can learn about the mergers debris structure in the action space.

This section only analyses the M31 galaxy in the 09-18~M31 simulation; however, we  tested that the results are similar to other M31/MW HESTIA galaxies. For reference, in Fig.~\ref{fig2::actions1}~(top row) we show the $\rm X-Z$ density distribution for the in situ stars and all accreted stars at $\rm z=0$ together with a population of surviving satellites~(different colours correspond to different dwarf galaxies). Since we focus on the stellar halo, we remove from our sample in situ and accreted stars located close to the galactic mid-plane~($|z|<1$~kpc) where the disc stars dominate and the impact of spiral arms and bar is more important in shaping the orbital action space~\citep{2017ApJ...839...61T,2021MNRAS.500.2645T}. 

In the second row of Fig.~\ref{fig2::actions1} we show the mass distribution for the in situ stars~(left), which are mostly confined near $(J_r,J_z)=(0,0)$ with a rather small tail of the distribution. Instead of showing the $(J_r,J_z)$ distribution for accreted stars, which do not demonstrate any prominent overdensities, we present the fraction of accreted stars in these coordinates. We see that accreted stars span over a larger area compared to the in situ stars, and only the accreted stars can be found at $(J_r,J_z)>1000\ {\rm kpc\ \kmps}$. In contrast to the smooth halo component, represented by the in situ and the mergers debris, despite some overlap, surviving satellites demonstrate a clear clustering in the $(J_r,J_z)$ space~(rightmost column in Fig.~\ref{fig2::actions1}). We notice the clustering of the satellites in other action spaces~(\JJtot and \sJrJp); however, this analysis does not deliver much information since the satellites are the distinct objects in galactic halo~(top right of Fig.~\ref{fig2::actions1}). 

Similar to $(J_r,J_z)$, we also do not detect any prominent substructures of the merger debris in the \JJtot or \sJrJp coordinates; however, the fractional distribution of accretes stars allows us to use the action space to disentangle between accreted and in situ stars that substantially overlap in \ELz and other phase-space coordinates~(see Sects.~\ref{sec2::assembly_history} and ~\ref{sec2::sausage}). This is better seen in the \sJrJp coordinates where stars above $\sqrt{J_r}>0.3-0.4\ {\rm (10^4\ kpc\ \kmps)^{0.5}}$ certainly represent accreted populations. Nevertheless, we suggest that the application of advanced clustering algorithms in the action space taking into account the chemical abundances of stars will make it possible to differentiate better the merger remnants.

\subsection{Individual merger debris in the action space}
In order to test how different mergers are mapped in the action spaces, in Fig.~\ref{fig2::actions2} we show the stellar mass distribution for five of the most significant mergers~(M1-M5) from the M31 galaxy~(09-18~simulation) in XZ, $(J_r,J_z)$, \JJtot and \sJrJp coordinates. There is not much diversity in the morphology of the merger remnants at $\rm z=0$ however, different mergers show a certain level of differentiation in $(J_r,J_z)$ space. One can notice that the M5 merger is elongated in the galactic plane having possibly larger radial (in cylindrical coordinates) motions. This is clearly seen in $(J_r,J_z)$ where the density distribution is stretched along the $J_r$ compared to the $J_z$. A more spherical remnant M3 has nearly equal distribution in both $J_r$ and $J_z$ actions. Finally, the M1 merger seems to be puffed up perpendicular to the plane and, as a result, its $(J_r,J_z)$ distribution is more elongated in $J_z$ compared to $J_r$. Since all the merger remnants slowly co-rotate with the galactic disc in \JJtot, all M1-M5 mergers mainly contribute to the right corner with complete overlap. A similarity in the net rotation is also evident from the \sJrJp plane where, except for the tails of the distributions, all the mergers occupy essentially the same region and are practically indistinguishable from each other. We note that the unmixed tidal tails of the merger debris can be captured in the action space, for example the M3 and M4 mergers, where some features outside the bulk of the distribution can be seen.

To summarize, the action space analysis of the HESTIA galaxies does not allow us to isolate accreted stars from the in situ in $(J_r,J_z)$, \JJtot and \sJrJp coordinates. However, some action ranges can be used to capture the accreted stellar populations. Different mergers debris show some systematic differences in the above-mentioned coordinates, despite the substantial overlap with each other. Since the dwarf galaxies are isolated substructures in the action space, the overlap between accreted populations suggests that the gravitational potential of the HESTIA galaxies grows non-adiabatically and/or dynamical friction significantly affects the trajectories of stars in the merger debris. Nevertheless, we suggest that some information from the actions still can be used to identify the merger debris.

We already noted some differences in the behaviour of the merger debris in $(J_r,J_z)$ space, where the ratio between vertical-to-radial actions correlates with the shape of the merger debris in the halo, which likely depends on the orbit of the infall and the internal structure of merging galaxies. For instance, for a disc-like component, one could expect that the distribution of stars is stretched along $J_r$, while a sphere should have a ratio similar to that of the $J_r, J_z$ actions. Therefore, we can introduce the angle between radial and vertical actions characterizing the shape of the debris $\Theta = atan(J_z/J_r)$. At the same time, stars from different debris could have different eccentricities in the galactic plane, also depending on the merger parameters. In Fig.~\ref{fig2::actions_ecc} we combine these two parameters and present the distribution of all stars~(top row), in situ~(second row), and all accreted stars~(third row) together with five individual merger debris with a higher stellar mass ratio relative to the main progenitor at the time of accretion~(M1-M5). In order to separate the prograde from the retrograde stars, we multiply the eccentricity by the sign of the angular momentum. The figure shows the density distribution of stars located in the galactocentric range of $(1-2)R_d$. Figure~\ref{fig2::actions_ecc} reveals many substructures where the in situ stars tend to have lower $\Theta$ and rather low eccentricities~(mainly $<0.5$), which is a typical behaviour of a regularly rotating disc-like component. Meanwhile, the accreted component shows a diverse morphology with some isolated clumps or overdensities with nearly constant or slightly varying eccentricity. The most important feature here is that in many cases we can see very different features typical for different merger debris, which, especially in combination with some other parameters or chemical abundances, can be used to clean the samples of accreted stellar populations. 

\subsection{Merger remnants: Combining orbital parameters and actions}
A number of recent studies suggest that highly eccentric orbits characterize the stars in the accreted MW halo population ~($>0.85$)~\citep[see, e.g.][]{2018ApJ...862L...1D,2018Natur.563...85H,2020MNRAS.494.3880B}. This result is supported by the analysis of the EAGLE simulations, which shows that dwarf galaxies with mass in the range $10^{8.5}-10^9$~\Msun accreted around $z\approx1.5$ would result in similar orbital parameters of the merger debris~\citep{2019MNRAS.482.3426M}. In Fig.~\ref{fig2::actions_ecc} we can see that the debris have a broad range of eccentricities, where the mean value could be different for different mergers. Since the EAGLE simulations include all types of galaxies with a small fraction of the MW-type objects~\citep{2018MNRAS.477.5072M}, it is worth investigating the eccentricity distribution of accreted stars in the HESTIA simulations, which were tailored to the LG galaxies.

In Fig.~\ref{fig2::eccentricity}~(left), we show the mean eccentricity of stars accreted from different satellite galaxies as a function of the accretion time and redshift. In the figure, we present all the mergers in all the M31/MW HESTIA galaxies colour-coded by the stellar mass of the merger. Similar to \cite{2019MNRAS.482.3426M} we find that more recent mergers have higher masses and result in the highly eccentric orbits of stars in the debris. It is worth noticing that the mean eccentricity values are indeed higher than $\approx0.6$. The left panel gives a somewhat misleading impression, suggesting that the merger debris result only in high-eccentricity stars. This becomes evident once we consider the distribution of eccentricities for all the stars in the mergers debris. In the right panel of Fig.~\ref{fig2::eccentricity}  we present the distribution of the eccentricity for the individual mergers debris~(grey) where the sign of the eccentricity is negative for stars with the negative angular momentum. We find that overall~(red lines) about $10-30\%$ of accreted stars have eccentricities below $0.5$. This result shows that the accreted stars do not necessarily have high eccentricities, and more recent events are more likely to produce populations with a disc-like kinematics.

\section{Dual nature of kinematically-defined stellar halo: In situ versus accreted populations}\label{sec2::dualism}

In this section we discuss the relations between in situ and accreted components of the stellar haloes in the HESTIA galaxies. In particular, we focus our analysis on the kinematically defined stellar haloes, which is done following a number of similar studies~\citep[see, e.g.][]{2010A&A...511L..10N,2018Natur.563...85H,2019A&A...631L...9K}, by using a Toomre diagram. In Fig.~\ref{fig2::toomre_diagram} we show the Toomre diagram for all stars~(top row), in situ stars only~(second row), and all stars from the merger remnants~(third row) at $\rm z=0$. To enhance the structure of the distributions in Fig.~\ref{fig2::toomre_diagram} we only consider stars located in $(1-3)~R_d$, which is roughly accessible for the MW large-scale surveys. In all the galaxies in our sample, the distributions on top show the presence of two distinct components, where one is centred near the LSR~($V_\phi \approx 200$~\kmps, disc). In contrast, the others tend to show slow prograde to retrograde rotation with a higher peculiar velocity. The latter one is mainly made of stars from disrupted dwarf galaxies -- mergers, presented in the third row. The properties and formation paths of the in situ stellar halo components are discussed in \citetalias{KhoperskovHESTIA-1}. 

To define the stellar halo in the Toomre diagrams we introduce a threshold separating stars with a halo-like kinematics from stars with a disc-like kinematics. We estimate the LSR velocity value $\rm V_{LSR}$ for each disc galaxy as the mean circular velocity value in the $(1-3)~R_d$ radial range. For the kinematically defined stellar halo we require a peculiar velocity component $\rm V_{pec} = \sqrt{V^2_R + V^2_z + (V_\phi - V_{LSR})^2}$ to be larger than $\rm V_{LSR} \times 180 \kmps/V^{MW}_{LSR}$, where $V^{MW}_{LSR}=240$~\kmps and $180$~\kmps are the values adopted in the MW to separate the disc from the halo~\citep[see, e.g.][]{2019A&A...632A...4D}. The boundary between the halo and disc kinematics is shown by the white lines in all the panels of Fig.~\ref{fig2::toomre_diagram}. Similar to a number of phase-space coordinates presented above, in Fig.~\ref{fig2::toomre_diagram} the individual mergers debris (fifth to eighth rows) have a diverse morphology, but interestingly most of them show a significant contribution to the region usually associated with the disc inside the white circle. We recall that we present only the structure of the most significant mergers while some others are missing, but their total contribution can be inferred from the third row. The structure of the Toomre diagrams in Fig.~\ref{fig2::toomre_diagram} shows that in a given radial range~$(1-3)R_d$ many accreted stars~($\approx 17-40\%$) contribute to the kinematically selected disc and, in contrast, the in situ populations~($20-50\%$) can be identified as the stellar halo. We suppose that these numbers are relevant for the MW. In that case, more work should be done in searching for the merger debris in the disc-like region of the Toomre diagram, also because in some cases, the individual merger debris is only deposited to this region~(e.g., M3 in the MW~09-18, M2 in the M31~17-11, M5 in the~17-11~MW and so on).

Figure~\ref{fig2::toomre_diagram} nicely illustrates the dual nature of the kinematically defined stellar halo, where it is made of both accreted and in situ stellar populations, demonstrating the complexity of the stellar halo and how its components overlap and affect each other. We already showed that in most of the HESTIA galaxies, there are only a few massive mergers in each galaxy~(see Fig.~4 in \citetalias{KhoperskovHESTIA-1}). At the same time, it is believed that the GSE merger is the biggest accretion event over the lifetime of the MW~\citep{2018MNRAS.478..611B,2018Natur.563...85H}. Therefore, we can test how much the most massive merger contributes to the accreted component of the kinematically defined stellar halo. In Fig.~\ref{fig2::masses}~(top) we show the fraction of the most massive merger~($M_{mmm}$) among all accreted stars as a function of the most massive merger mass. The figure shows that $20-70\%$ of the accreted mass in the M31/MW-type disc galaxies comes from a single merger with a median value of about $40\%$ and the fraction of a single merger is larger for more massive accretion events. We note that the correlation does not depend on the definition of the accreted component (either total or kinematically defined halo). 

Finally, in the bottom panel of Fig.~\ref{fig2::masses} we show how the fraction of the in situ kinematically defined stellar halo depends on the total accreted stellar mass. With a single exception, the MW in the 37-11 simulation, we find a positive correlation, which is somewhat counter-intuitive because, in this case, the fraction of accreted halo components is lower for more massive ex situ haloes. Therefore, more accreted material results in a more heavily perturbed disc, which deposits more stars to the kinematically defined halo. Since these relations are based on the constrained simulations of the LG, they can be used to estimate the total accreted mass once the in situ halo is constrained, for example by using chemical abundances.

\section{Summary}\label{sec2::concl}
In this work we analysed six M31/MW analogues in the HESTIA suite of cosmological hydrodynamics zoom-in simulations. All the galaxies experienced between one and four significant mergers with stellar mass ratios between 0.2 and 1~(relative to the host), where all the significant mergers~(with a single exception) happened between 7 and 11~Gyr ago. We studied the merger debris using integrals of motions~(total energy and angular momentum), phase-space coordinates, orbital parameters, and actions. Focusing our analysis on the most significant merger debris, with the larger stellar mass fraction relative to the host galaxy at the time of accretion, we arrived at the following conclusions:

\begin{itemize}
    \item Different stellar merger debris show a diverse morphology at the time of accretion, varying from streams and shell-like structures to smooth nearly spherical distributions~(see Fig.~\ref{fig2::mergers_initial}). At redshift zero, the vast majority of stellar merger debris have similar and almost featureless density distributions, flattened towards the disc plane. Some exceptions were found for the most recent mergers, which show some unmixed substructures~(see Fig.~\ref{fig2::mergers_final}). More ancient merger debris are more compact compared to more recent accretion events, likely due to the mass growth of the galaxy over time, deepening the potential well.
    
    \item  The self-consistent cosmological HESTIA simulations suggest the \ELz distributions of individual merger debris evolve in time due to the host's mass growth and its non-axisymmetric gravitational potential~(see Fig.~\ref{fig2::e_lz_all}). Therefore, we demonstrate that neither energy nor angular momentum of the merger remnants are conserved after the merger has concluded. In some cases, we observe a substantial change in the direction of the angular momentum and noticeable changes in the energy distribution of the stellar debris.
    
    \item We showed that merger remnants from a single event are not clustered in the integrals of motion space, but cover a large area in these coordinates. At the same time, however, the \ELz distributions are not featureless as a single merger remnant often contributes to a few overdensities~(see Fig.~\ref{fig2::e_lz_all}).
    
    \item By dissecting the \Elz distributions by age and the location inside dwarf galaxies before the mergers, we found that inside the host galaxy, the accreted stars with lowest total energy are typically older and were acquired from the outer parts of the dwarf galaxies~(see Fig.~\ref{fig2::ages_of_mergers}). This variation in parameters is translated into a prominent radial positive age gradient for individual merger debris. However, the mean age of the debris is largely biased towards the time of accretion because most of the mergers are accompanied by an enhancement of star formation. Moreover, the most significant bursts of star formation inside the merger debris correspond to the merger time. This result also suggests that in the MW the merger time estimation can be done by using the age distribution of the stellar debris, but that the accreted stars from the innermost disc should be taken into account. 

    \item The merger remnants show a diverse appearance in phase-space coordinates with some examples similar to the ones discovered in the \Gaia data. In particular, the HESTIA simulations reproduce the Gaia-Sausage-like features in the \UV plane with prominent non-rotating or weakly counter-rotating components~(see Fig.~\ref{fig2::uv}). Moreover, in some cases we can see the presence of more that one Gaia-Sausage-like feature~(see Fig.~\ref{fig2::sausage_evolution}). We also demonstrate that the local \UV Gaia-Sausage-like features are the parts of large-scale radially extended components traceable across the entire disc~(see Fig.~\ref{fig2::r_vphi}).  
    
    \item Apart from the dwarf galaxy population, the merger debris do not show any isolated structures in the action space being distributed over a large area, as well as overlapping with each other and in situ stars~(see Figs.~\ref{fig2::actions1},~\ref{fig2::actions2}). However, we suggest that accreted stars can more likely be identified with the cut $\sqrt{J_r}>0.2-0.3\ {\rm (10^4\ kpc\ km\ s^{-1})^{0.5}}$. We also propose that some merger debris can be disentangled from each other in the ($J_z/J_r$ -- orbital eccentricity) space. Nevertheless, we showed that the actions, once combined with the orbital parameters~(e.g. eccentricity), demonstrate different behaviour for different merger debris, thus making it possible to disentangle accretion events~(see Fig.~\ref{fig2::actions_ecc}). 
    
    \item  The accreted stars in HESTIA have a broad distribution of eccentricities, peaking at $\approx0.6-0.9$. The mean eccentricity of the accreted stars correlates with the mass of the accreted dwarf galaxy and the time of accretion. In particular, more massive mergers and more recent ones result in lower eccentricity of stars at $\rm z=0$. 
    
    \item Finally, we showed that identifying stellar haloes in the Toomre diagram selects both accreted and in situ stars, which in total represent $45-65\%$ of the stellar mass of the galaxy~(excluding dwarf galaxies). The accreted component of such kinematically defined stellar haloes is mainly made of a single merger~($\approx 40-50\%$). The in situ mass fraction also correlates with the total accreted mass~(see Fig.~\ref{fig2::masses}). 

\end{itemize}

The structure of accreted stellar populations in the MW provides valuable insights into its assembly history. Accreted stars typically exhibit distinct kinematic and chemical properties compared to the stars that formed in situ. These accreted populations often display more random and retrograde orbits, indicating their external origin. In the HESTIA galaxies, we showed how the accreted stars contribute to the growth of stellar halos and analysed kinematic signatures of past mergers events. Our study demonstrates how the structure of accreted stellar populations helps unravel the dynamical processes and interactions that have shaped the MW galaxy over cosmic time.

\begin{acknowledgements}
We thank the anonymous referee for their valuable comments. SK acknowledgements the HESTIA collaboration for providing access to the simulations.
FAG acknowledges support from ANID FONDECYT Regular 1211370 and from the ANID BASAL project FB210003. FAG,  acknowledge funding from the Max Planck Society through a Partner Group grant. AK is supported by the Ministerio de Ciencia e Innovaci\'{o}n (MICINN) under research grant PID2021-122603NB-C21 and further thanks Smashing Pumpkins for Siamese dreams. JS acknowledges support from the French Agence Nationale de la Recherche for the LOCALIZATION project under grant agreements ANR-21-CE31-0019. YH has been partially supported by the Israel Science Foundation grant ISF 1358/18.
ET acknowledges support by ETAg grant PRG1006 and by EU through the ERDF CoE TK133.

\newline
{\it Software:} \texttt{IPython} \citep{2007CSE.....9c..21P}, \texttt{Astropy} \citep{2013A&A...558A..33A, 2018AJ....156..123A}, \texttt{NumPy} \citep{2011CSE....13b..22V}, \texttt{SciPy} \citep{2020SciPy-NMeth}, \texttt{AGAMA} \citep{2019MNRAS.482.1525V}, \texttt{Matplotlib} \citep{2007CSE.....9...90H}, \texttt{Pandas} \citep{mckinney-proc-scipy-2010}, TOPCAT~\citep{2005ASPC..347...29T}.
\end{acknowledgements}

\bibliographystyle{aa}
\bibliography{references-1}

\end{document}